\documentclass[aps,prl,twocolumn,superscriptaddress,amsmath,amssymb,aps]{revtex4-2}

\usepackage[hidelinks]{hyperref}

\usepackage[colorinlistoftodos]{todonotes}
\usepackage{graphicx}
\usepackage{dcolumn}
\usepackage{subcaption}
\usepackage{esint}
\usepackage{amssymb}
\usepackage{bm}
\usepackage{xcolor}
\usepackage{listings}
\usepackage[normalem]{ulem}

\newcommand{\Ree}{\mathfrak{Re}}
\newcommand{\Imm}{\mathfrak{Im}}

\begin{document}

\title{Random initial data and average shock time in the Fermi-Pasta-Ulam-Tsingou chain}







%

\author{Matteo Gallone}
\email{mgallone@sissa.it}
\affiliation{Scuola Internazionale di Studi Superiori Avanzati, Via Bonomea 265, 34136 Trieste, Italy} 

\author{Ricardo Grande}
\email{rgrandei@sissa.it}
\affiliation{Scuola Internazionale di Studi Superiori Avanzati, Via Bonomea 265, 34136 Trieste, Italy}

\author{Antonio Ponno}
 \email{ponno@math.unipd.it}
\affiliation{Department of Mathematics ``T. Levi-Civita'', University of Padova, Via Trieste 63, 35131 Padova, Italy}

\author{Stefano Ruffo}%
 \email{ruffo@sissa.it}
\affiliation{ISC-CNR, via Madonna del Piano 10, 50019 Sesto Fiorentino, Firenze, Italy}
 \affiliation{Scuola Internazionale di Studi Superiori Avanzati, Via Bonomea 265, 34136 Trieste, Italy}
\address{INFN, Sezione di Firenze, 50019 Sesto Fiorentino, Italy }

\author{Erwan Druais}
\email{erwan.druais@ens-lyon.fr}
\affiliation{\'Ecole Normale Sup\'erieure de Lyon, 15 parvis René Descartes, 69342 Lyon, France}

\date{\today}

\begin{abstract}
We investigate the dynamics of the Fermi–Pasta–Ulam–Tsingou chain with long-wavelength random initial data. When the energy per particle is small, thermal equilibrium is not reached on a fast timescale and the system enters prethermalization. The formation of the prethermal state is characterized by the development of a Burgers-type shock and the onset of a turbulent-like spectrum with a time dependent exponent $\zeta(t)$ in the inertial range. We perform a significant step forward by demonstrating that these features are robust under generic long-wavelength random initial conditions. By employing advanced probabilistic techniques inspired by the works of Dudley and Talagrand, we derive a sharp asymptotic expression for the average shock time in the thermodynamic limit. For large $p$, this time scales as $(p \sqrt{\log p})^{-1}$, where $p$ is the number of excited modes proving that it is an intensive quantity up to a logarithmic correction in the size of the system.
\end{abstract}

\maketitle

\textit{Introduction.---} Understanding how isolated physical systems approach thermal equilibrium is a central open problem in statistical mechanics. The microscopic mechanisms by which macroscopic systems redistribute energy among their degrees of freedom are complex and not yet fully understood \cite{Balescu1997-xs}. This complexity was brought to light by the numerical simulations of Fermi, Pasta, Ulam and Tsingou (FPUT) \cite{FPU55}. In their pioneering work, they integrated numerically a one-dimensional chain of particles and followed the time evolution of the Fourier Energy Spectrum (FES). Rather than reaching equipartition, which is a necessary condition for \emph{thermal equilibrium}, the system exhibited unexpected recurrent dynamics.
Since then, FPUT-like recurrences have been reported across a wide range of physical settings, including holography \cite{Balasubramanian2014}, graphene resonators \cite{Midtvedt2014}, nonlinear phononic lattices \cite{Cao2014}, photonic systems \cite{Pierangeli2018}, trapped cold atoms \cite{Kinoshita2006,Villain2000,Danshita2014}. 

Random initial data drawn from the Gibbs measure have played a crucial role in the study of the dynamics of coupled harmonic oscillators \cite{Ford1965}. Initial conditions close to thermal equilibrium have been employed when studying anomalous conduction in one-dimensional systems of coupled nonlinear oscillators \cite{Lepri1997,Prosen2000,Lepri2003,Mendl-Spohn-2013}. With a different perspective, the robustness of dynamical behavior with respect to random perturbations of the invariant measure has been examined \cite{Young1986,Kifer1988,AST-2003-286-25-0}. 
 By contrast, initial data far from statistical equilibrium or systems subject to strong random perturbations have more rarely been considered. These regimes, which capture how typical nonequilibrium configurations relax toward thermal equilibrium, are the focus of this Letter.

\begin{figure}[h!]
	\includegraphics[width=0.4\textwidth]{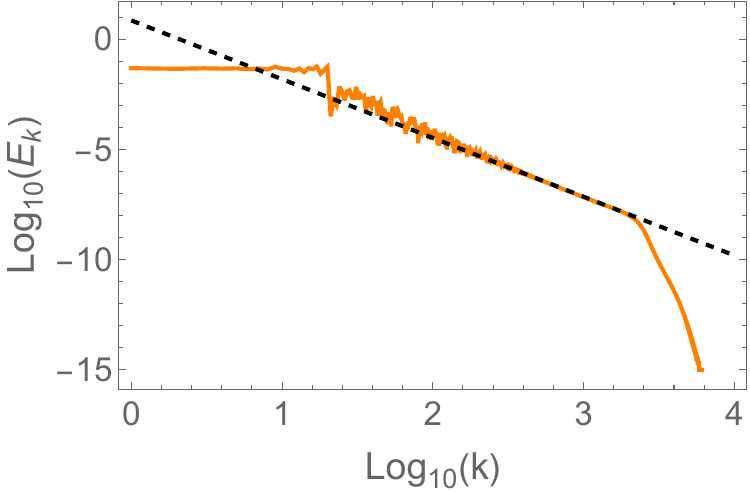}
	\caption{Fourier Energy Spectrum (full orange line) at the shock time with $\epsilon=10^{-3}$ for $\alpha=1$, $\beta=0.1$ and $N=16\,384$. $20$ modes are initially excited. The black dashed line is the theoretical prediction $\log E_k =-8/3 \log k + 0.86$.}\label{FigSpectro}
\end{figure}

Concerning the time evolution of the FPUT system, numerical studies have shown that, when energy is above a certain threshold, the system undergoes a quick thermalization and the FPUT recurrence is not present \cite{Chirikov1979}. The process by which a system reaches thermal equilibrium is called \emph{thermalization}~\cite{Mori2018-pk}.  While approaching thermal equilibrium, the system may become trapped for a long time in a quasi-stationary \emph{prethermal} state. Indeed, this is what happens in the FPUT system when the initial energy is below the threshold mentioned above~\cite{GallavottiBook2008}. 
This initial stage, known as \emph{prethermalization}, has attracted broad interest beyond the FPUT dynamics, including quantum systems \cite{Berges2004,Luitz2020,Kyprianidis2021-gu,Stasiuk2023}, because of its deep experimental and technological relevance \cite{Mori2018-pk,Oka2019,He2023}.  We emphasize that prethermalization corresponds to the relaxation to a quasi-stationary state different from thermal equilibrium, preceding the eventual onset of thermal equilibrium on longer time scales~\cite{Mori2018-pk}.
In the FPUT system, prethermalization is strongly dependent on the choice of the initial condition.  Numerical evidence and heuristic arguments suggest that, for certain classes of out-of-equilibrium random initial data, prethermalization is governed solely by intensive parameters \cite{Livi1985,Kantz1994,Benettin2008-1D}.  We will indeed show that, up to a logarithmic correction, the onset of prethermalization in the FPUT system is uniquely governed by the specific energy and the fraction of initially excited modes with random phases.

 Prior to tackling the case of random initial conditions, a transient turbulent regime was identified in a recent work for single-mode initial conditions
\cite{Gallone2022-PRL,Gallone2024}. This regime  highlights a quantitative mechanism for the redistribution of energy from low to high modes of the Fourier spectrum. This regime is present for large wavelength initial conditions and at sufficiently short timescales. 
The key idea was to show that, on such a timescale, the FPUT dynamics can be well approximated by the inviscid Burgers equation. 
Solutions of the Burgers equation exist up to a finite time $t_s$, the shock time. 
At $t_s$, the FPUT displays a turbulent-like FES with an inertial-range scaling $E_k \sim k^{-8/3}$.
Thus, the shock time emerged as the fundamental timescale for the formation of the prethermal state in the FPUT chain \cite{Gallone2022-PRL}.

In this Letter we show analytically and confirm numerically that the Burgers turbulence scenario is robust for out-of-equilibrium long-wavelength random initial data (see Fig.~\ref{FigSpectro}).
Developing advanced probabilistic techniques inspired by the works of Dudley and Talagrand \cite{Dudley1967,Talagrand2014}, we derive an explicit analytic formula for the average shock time in the thermodynamic limit for a finite and large fraction of excited modes. 
The theory rigorously explains the observed differences in scaling behavior between \emph{coherent} initial data where the phases of the Fourier modes are all equal or equispaced, and \emph{incoherent} initial data where the phases are taken randomly on the unit circle \cite{Benettin2008-1D}.
In addition to the power-law scaling, we analytically predict a logarithmic correction, which we confirm numerically (Fig.~\ref{FigLog}).


\textit{Model, initial conditions and main result.---} The FPUT model consists in $N$ masses interacting nonlinearly on a one-dimensional lattice. The Hamiltonian is
\begin{equation}\label{eq:FPUT}
	H=\sum_{j=1}^N \left[\frac{p_j^2}{2}+V(q_{j+1}-q_j) \right] \, ,
\end{equation}
where $V(z)=z^2/2+\alpha z^3/3+\beta z^4/4$, $q_j$ is the displacement from equilibrium of the $j$-th mass, and $p_j$ its momentum and, without loss of generality, we take the masses and the harmonic constant equal to one and we consider periodic boundary conditions $p_{N+1}=p_1$, $q_{N+1}=q_1$.

When $\alpha=\beta=0$, the system is a linear harmonic chain which is completely integrable. The dynamics of the FES becomes trivial, and no exchange of energy among normal modes is allowed. When $\alpha\neq 0$ or $\beta\neq0$, the nonlinearity drives thermalization by coupling the modes and allowing for energy exchange. Although the latter is very common, the nonlinearity by itself is not sufficient to drive thermalization \cite{Toda1967-B,Hnon1974,Goldfriend2019}. 

The energy of the mode $k$ is defined as 
\begin{equation}
	E_k(t)=\frac{|\hat{p}_k(t)|^2}{2}+\frac{\omega_k^2 |\hat{q}_k(t)|^2}{2} \, ,
\end{equation} 
where $\omega_k=|2 \sin( \pi k/N)|$, $\hat{q}_k$ and $\hat{p}_k$ are the Fourier coefficients, e.g.~$\hat{q}_k =(1/\sqrt{N}) \sum_{j=1}^N q_j e^{ i 2 \pi k j/N}$. We consider the initial data
\begin{eqnarray}\label{eq:InitialCondi}
	q_j(0)&=&\sum_{k=1}^p \frac{A_k}{\sqrt{p}} \sin(\psi_k) \cos\left(\frac{2 \pi k j}{N}+\phi_k\right) \, , \\
	p_j(0)&=&\sum_{k=1}^p \frac{\omega_k A_k}{\sqrt{p}}\cos(\psi_k) \sin\left(\frac{2 \pi j k}{N}+\theta_k\right) \, ,\label{eq:InitialCondi2}
\end{eqnarray}
where $A_k$ is the amplitude of the $k$-th mode, and $\psi_k,\phi_k,\theta_k$ are the phases. The ratio between kinetic and potential energy in the mode $k$ is determined by $\psi_k$, while $\phi_k$ and $\theta_k$ are relative spatial offsets between the normal modes.

\begin{figure}[h!]
	\includegraphics[width=0.4\textwidth]{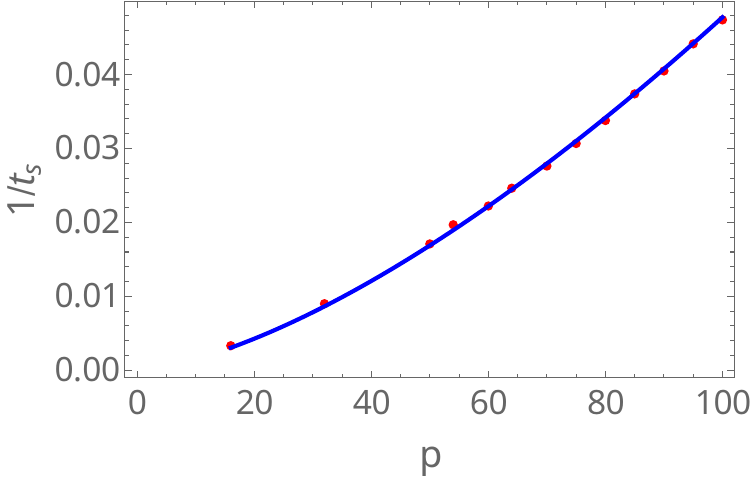}
	\caption{Numerical estimate of the inverse shock time as a function of the number of excited modes $p$ with all phases equal to zero (points), eq.~\eqref{eq:InitialCondi}, \eqref{eq:InitialCondi2}. The solid blue line is the interpolation curve $4.774 \times 10^{-5} p^{\frac{3}{2}}$. $\alpha=1$, $\beta=0.1$, $\epsilon=10^{-3}$, $N=2048$.} \label{Figp32}
\end{figure}

\begin{figure}[h!]
	\includegraphics[width=0.4\textwidth]{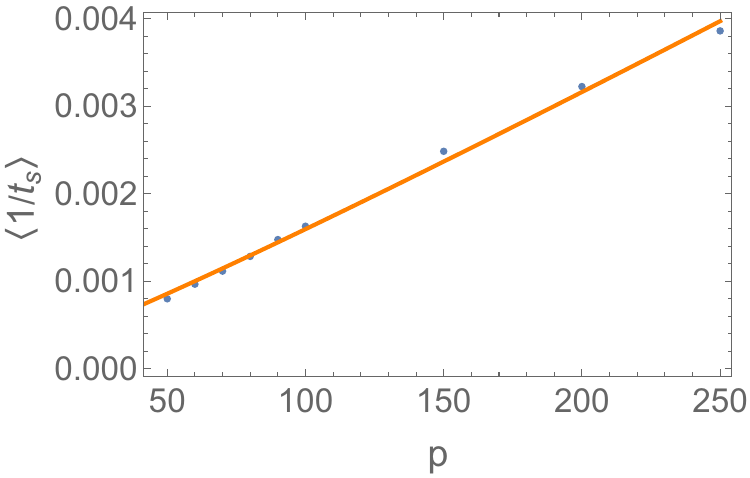}
	\caption{Average of the inverse shock time $\langle t_s^{-1} \rangle$ vs. the number of initially excited modes $p$ (points) eq.~\eqref{eq:InitialCondi}, \eqref{eq:InitialCondi2}. Each point is the average over $1\,000$ realizations of the phases. The solid orange line is the  theoretical interpolation curve $p \sqrt{\log p}$. $\alpha=1$, $\beta=0.1$, $\epsilon=10^{-3}$, $N=16\,384$. }\label{FigLog}
\end{figure}

We consider initial conditions where $E_k(0)=\epsilon/p$ for $1\leq k \leq p$ and $E_k(0)=0$ otherwise. We then prove the existence of a timescale $t_s$ at which the FES has a window which exhibits a power-law spectrum
\begin{equation}
	E_k(t_s) \sim k^{-\frac{8}{3}} \, , \quad k_1 \leq k \leq k_c \, .
\end{equation}
The lower bound $k_1$ is empirically determined, while $k_c$ is placed at the beginning of the exponential cutoff of the spectrum. 
The width of the window is intensive in $N$ and $t_s$ and scales with the number of excited modes $p$ depending on the relation between phases. When the phases $\psi_k,\theta_k,\phi_k$ are coherent (e.g.~all vanishing or all equispaced), then $t_s$ is a deterministic variable and we find that $t_s^{-1} \sim \frac{\alpha \sqrt{\epsilon}}{N} p^{\frac{3}{2}}$ (see Fig.~\ref{Figp32}). If the phases are chosen randomly, then $t_s$ becomes a random variable and we show that $t_s$ is much longer on average, with
\begin{equation}\label{eq:Supertramp}
	\langle t_s^{-1} \rangle \sim \frac{\alpha \sqrt{\epsilon} \, p \sqrt{\log p}}{N} \, ,
\end{equation}
for large $p$ (see Fig.~\ref{FigLog}). Here $\langle \, \rangle$ denotes the average over the phases.

When a finite small fraction $\nu$ of modes is initially excited, i.e.~$p=\nu N$, then \eqref{eq:Supertramp} yields $\langle t_s^{-1} \rangle \sim \alpha \sqrt{\epsilon} \nu \sqrt{\log N}$ which means that, apart from a logarithmic correction $\sqrt{\log N}$, $\langle t_s^{-1} \rangle$ is ruled by intensive parameters only.

\textit{Continuum approximation.---} In order to study the evolution of the initial condition \eqref{eq:InitialCondi} for small specific energy and large number of particles we introduce the two fields $Q(x,\tau)$ and $P(x,\tau)$ of spatial period one, such that $q_j(t)=NQ(j/N,t/N)$, $p_j(t)=P(j/N,t/N)$. Following \cite{Gallone2022-PRL}, we note that $\epsilon=\frac{1}{N}\sum_{k=1}^p E_k(0)$, up to higher order corrections in $\epsilon$. We introduce the Left- and Right- fields $L=(Q_x+P)/(\sqrt{2 \epsilon} )$ and $R=(Q_x-P)/(\sqrt{2 \epsilon})$, where partial derivatives are denoted by subscripts. Neglecting higher order terms in $\epsilon$ and $\frac{1}{N^2}$, the evolution equations are
\begin{eqnarray}
	L_\tau&=&\Big( L+\frac{\alpha \sqrt{\epsilon}}{2 \sqrt{2}} (L+R)^2 \Big)_x \, ,\\
	R_\tau&=&-\Big(R+\frac{\alpha\sqrt{\epsilon}}{2 \sqrt{2}} (L+R)^2 \Big)_x \, .
\end{eqnarray}
Since the equations for $L$ and $R$ are nonlinearly coupled, their analysis is simplified by means of a close-to-identity canonical transformation $\mathcal{C}_{\sqrt{\epsilon}}:(L,R) \mapsto (\lambda,\rho)$ which decouples the equations for the new fields $\lambda$ and $\rho$ at order $\sqrt{\epsilon}$. Neglecting terms of order $\epsilon$, the transformed equations are
\begin{equation}\label{eq:Burgerissime}
	\lambda_\tau=\Big(1+\frac{\alpha\sqrt{\epsilon}}{\sqrt{2}} \lambda \Big)\lambda_x \, , \quad \rho_\tau=-\Big(1+\frac{\alpha \sqrt{\epsilon}}{\sqrt{2}} \rho \Big)\rho _x
\end{equation}
with initial conditions $(\lambda(x,0),\rho(x,0))=(L(x,0),R(x,0))+O(\sqrt{\epsilon})$. Equations \eqref{eq:Burgerissime} are a couple of inviscid Burgers equations, whose solution is well-defined for times $\tau< \tau_s$ where $\tau_s$ is the scaled shock time of the system, i.e.~$\tau_s=\min\{\tau_s^\rho,\tau_s^\lambda\}$ and $t_s=N\tau_s$.

\textit{Statistics of the shock time.---} Using Formula (7) and Section V-A in \cite{Gallone2024}, the large $k$ asymptotics of the FES at $\tau_s$ for initial data of the form \eqref{eq:InitialCondi} and \eqref{eq:InitialCondi2} is $E_k(\tau_s)\sim k^{-\frac{8}{3}}$.

The scaled time $\tau_s^\eta$ is given by
\begin{equation}\label{eq:Canutazzu}
	\begin{split}(\tau_s^\eta)^{-1}&=\frac{\alpha \sqrt{\epsilon}}{\sqrt{2}} \max_{x \in [0,1]} \eta_x(x) \, , \\ \eta(x)&=\lambda(x,0),\rho(x,0) \, .
	\end{split}
\end{equation}
Using this formula, the computation of the shock time reduces to a maximization problem for each realization of the initial data.
For coherent initial phases in \eqref{eq:InitialCondi} and \eqref{eq:InitialCondi2} (e.g.~all equal to zero), the maximization reduces to finding the maximum of $1/\sqrt{p}\sum_{k=1}^p 2 \pi k \cos(2 \pi k x)$ which occurs at $x=0$ and scales asymptotically as $p^{\frac{3}{2}}$ for large $p$.
For random initial phases, $\tau_s$ becomes a random variable and computing its average reduces to studying the extremal statistics of the stochastic process $\eta_x(x)$. 
We show that, on average, the scaled shock time is significantly longer than in the coherent case, with$\langle \tau_{s}^{-1} \rangle \sim \alpha \sqrt{\epsilon} p \sqrt{\log p}$.
Further details on the argument are given in the supplemental material \cite{supplemental}, while a streamlined version is presented below.




Since the maximum in Eq.~\eqref{eq:Canutazzu} is taken over infinitely many points, we first restrict to a finite set, for which the statistics can be bounded as:
\begin{equation}\label{eq:union_bound}
	\begin{split}
	\Big\langle&\max_{m=1,\dots,M}\eta_x(x_m)\Big\rangle  
    \\&= \int_0^{+\infty} \mathrm{Prob}\Big(\max_{m=1,\dots,M} \eta_x(x_m)) >\lambda\Big) \, d \lambda\\
	&\leq M \int_0^{+\infty} \mathrm{Prob}\Big(\eta_x(x_1)>\lambda\Big) \, d \lambda \, .
	\end{split}
\end{equation}
Because the maximum we want to evaluate is taken over an infinite set, these simple bounds diverge as $M\to \infty$. However, $\eta_x$ is a smooth function, so its values at nearby points are not statistically independent. These correlations can thus be exploited to improve the bound. In particular, we are able to derive the following sub-Gaussian bound
\begin{equation}\label{tailbound}
	\mathrm{Prob}(|\eta_x(x)-\eta_x(y)| > \lambda) \lesssim \exp\left(-\frac{\lambda^2\, p}{4 d_p(x,y)^2}\right)
\end{equation}
where $d_p(x,y)^2=p\,\langle|\eta_x(x)-\eta_x(y)|^2\rangle$ is the desired measure of correlation. For our process, we derive
\begin{equation}\label{dp}
    d_p(x,y)^2=\frac{1}{2} \sum_{k=1}^p k^2\left(1-\cos\big(2 \pi k (x-y)\big)\right) \, .
\end{equation}

In order to carefully combine the bounds \eqref{eq:union_bound} and exploit correlations, we introduce a family of point lattices $\mathcal{N}_m$ such that, for any $m'<m$, $\mathcal{N}_{m'}$ forms a sublattice of $\mathcal{N}_m$. The optimal way to create this sublattice is to select points which are equally spaced with respect to the distance $d_p$. The mesh size $r^m$ must decrease exponentially fast, i.e.~$r^m$ for $r\in (0,\frac18)$ sufficiently small.

The next step consists in a chaining argument: for any $x \in [0,1)$ we construct a sequence of points $x_m \in \mathcal{N}_m$ converging monotonically to $x$ and such that
\begin{equation}
	\eta_x(x)=\eta_x(0) + \sum_{m=1}^{\infty} [\eta_x(x_m)-\eta_x(x_{m-1})] \, .
\end{equation}
Then, taking the max on both sides, one finds
\begin{equation}
	\max_{x \in [0,1]} |\eta_x(x)-\eta_x(0)|\leq \sum_{m=1}^{\infty}\max_{x \in [0,1]} |\eta_x(x_m)-\eta_x(x_{m-1})| \, .
\end{equation}
For each fixed $m$, the sum involves a maximum over a finite set of points with cardinality $ |\mathcal{N}_m| |\mathcal{N}_{m-1}|$, which allows the application of \eqref{eq:union_bound}. We then define $\sigma_m=(8r)^m \sqrt{\log(2^m|\mathcal{N}_m| |\mathcal{N}_{m-1}|)}$ and $\sigma=\sum_{m=1}^\infty \sigma_m$. We can then show that $\max_{x \in [0,1]} \eta_x(x)$ satisfies the following sub-Gaussian bound:
\begin{equation}
    \begin{split}
        &\text{Prob}\Big(\max_{x\in [0,1]} |\eta_x(x)-\eta_x(0)| > \frac{\lambda \sigma}{\sqrt{p}}\Big)  \\ &\leq \text{Prob}\Big(\sum_{m=1}^\infty |\eta_x(x_m)-\eta_x(x_{m-1})| > \frac{\lambda \sigma}{\sqrt{p}}\Big) \\
        &\lesssim \sum_{m=1}^\infty |\mathcal{N}_m| |\mathcal{N}_{m-1}| \exp\Big(-\frac{\lambda^2 \sigma_m^2}{8\,d_p(x_m,x_{m-1})^2}\Big)\\
        &\lesssim 2^{-\lambda^2} \, .
    \end{split}
\end{equation}
Using Fubini's theorem \cite{liebLoss}
\begin{equation}\label{eq:AlbertoPausa}
    \begin{split}
        &\langle \max_{x \in [0,1]} |\eta_x(x)-\eta_x(0)| \rangle \\
        &=\int_0^\infty \text{Prob}\Big(\max_{x \in [0,1]} |\eta_x(x)-\eta_x(0)| > \lambda\Big) d\lambda \\
        &=\frac{\sigma}{\sqrt{p}} \int_0^\infty \text{Prob}\Big(\max_{x \in [0,1]} |\eta_x(x)-\eta_x(0)| > \frac{\lambda \sigma}{\sqrt{p}}\Big) d \lambda \\&\leq \frac{\sigma}{\sqrt{p}} \, .
    \end{split}
\end{equation}
It thus remains to estimate $\sigma$, which can be bounded as follows:
\begin{equation}\label{eq:QUasiDudley}
    \sigma \lesssim \int_0^{+\infty} \sqrt{\log N_p(\varepsilon)} \, d \varepsilon
\end{equation}
where $N_p(\varepsilon)$ is the packing number associated with the distance $d_p$; that is, the maximum number of points in $[0,1]$ whose $d_p$-distance is at least $\varepsilon$. The main contribution to the integral comes from the small $\varepsilon$ region. A similar bound was established by Dudley for Gaussian processes \cite{Dudley1967}.
In order to compute the packing number, we estimate \eqref{dp} as follows:
\begin{equation}\label{eq:euclidean_vs_dp}
d_p(x,y)^2\leq 2\pi^2\, |x-y|^{(\log p)^{-1}}\, \sum_{k=1}^p k^{2+(\log p)^{-1}}.
\end{equation}
This quantitative comparison between the canonical distance $d_p$ and the Euclidean distance allows us to compare their respective packing numbers, the latter being explicitly computable.
This leads to an upper bound on $N_p(\varepsilon)$ which, when combined with \eqref{eq:QUasiDudley}, gives $\sigma \lesssim p^{\frac{3}{2}} \sqrt{\log p}$. Together, estimates \eqref{eq:Canutazzu} and \eqref{eq:AlbertoPausa}, imply an upper bound for the inverse of the shock time.

Concerning the lower bound, following the pioneering work of Talagrand \cite{Talagrand1996,Talagrand2014}, we show that
\begin{equation}\label{eq:Talagrand}
    \langle \max_{x \in [0,1]} \eta_x(x) \rangle \gtrsim \frac{1}{\sqrt{p}} \int_0^{+\infty} \sqrt{\log N_p(\varepsilon)} \, d\varepsilon \, .
\end{equation}
For $p\gg 1$ and $\varepsilon\ll p^{\frac{3}{2}}$, we explicitly construct $p^{\frac52} \varepsilon^{-1}$ points in $[0,1]$ whose distance is at least $\varepsilon$ when measured with respect to $d_p$, thereby realizing the packing number $N_p(\varepsilon)$.

The construction of this large set of $\varepsilon$-separated points relies on two key ideas. First, a local approximation $d_p(x,0)\sim p^{\frac{3}{2}}\, [px + o(px)]$ allows us to construct a cluster of $p^{\frac{3}{2}}\,\varepsilon^{-1}$ points in the interval $[0,\frac{1}{p}]$ that are $\varepsilon$-separated. Summing the series \eqref{dp}, one finds that $d_p(x,y)> p^{\frac{3}{2}}$ whenever $|x-y|\gtrsim p^{-1}$. 
This enables the construction of additional clusters of $p^{\frac{3}{2}}\,\varepsilon^{-1}$ points, provided they are $p^{-1}$-separated in the Euclidean distance. Altogether, this yields a set of $p^{\frac{5}{2}}\,\varepsilon^{-1}$ points which are $\varepsilon$-separated with respect to the distance $d_p$. This yields the lower bound $N_p(\varepsilon) \gtrsim p^{\frac52} \varepsilon^{-1}$, which, together with \eqref{eq:Canutazzu} and \eqref{eq:Talagrand}, provides a sharp lower bound for the inverse of the shock time.

\textit{Conclusions.---} In this Letter we have shown that the Burgers turbulence scenario in the FPUT model is robust when $p$ long-wavelength modes
are initially excited with random phases. By developing advanced probabilistic techniques based on \cite{Dudley1967,Talagrand2014}, we obtained an analytic expression for the large $p$ asymptotics of the average shock time.
This computation is crucial for understanding the formation of the prethermal state for typical initial conditions.

Our approach explains the previously observed differences in the scaling behavior of the prethermal state for incoherent versus coherent phases \cite{Benettin2008-1D}. Up to a logarithmic correction, when an extensive fraction of modes is initially excited, two distinct behaviors emerge: for random phases, the average shock time \eqref{eq:Supertramp} is governed solely by intensive parameters, whereas for coherent phases it vanishes as $N\to\infty$. This implies that the scenario in which the shock time vanishes corresponds to a measure-zero set of synchronized initial phases and is thus not relevant for the average from a statistical mechanics point of view. The derivation of the logarithmic corrections relies on sophisticated probabilistic techniques, detailed in the supplemental material \cite{supplemental}.

To place our results in context, we recall that the evolution of integrable systems with random initial data can mimic certain aspects of hydrodynamic turbulence, such as power-law spectra and the formation of coherent structures \cite{Zakharov2009,SotoCrespo2016,Agafontsev-Zakharov2020}. Within the approximation considered here, the dynamics under long-wavelength random initial data fits naturally within the framework of integrable turbulence.
In fact, our techniques also allow one to derive the full statistics of the shock time for the Burgers equation when random initial data are considered. 
The statistical properties of shocks in the Burgers equation with various classes of random initial data have been the subject of extensive study 
\cite{Kida1979,fournier1983equation,Sinai1992,She1992}.
Rigorous results on the spatial statistics of shocks in the zero-dissipation limit have also been obtained
\cite{Sinai1992,Avellaneda1995}. To our knowledge, however, the statistics of the first shock time --corresponding to the onset of the turbulence regime in FPUT-- has  not been previously studied. Addressing this question requires the technical apparatus developed in this Letter.

The discrepancy in the spatial profile between Burgers dynamics and lattice FPUT model arises shortly before the shock time. At this time, the dispersion of the lattice model is responsible for the generation of a train of short wavelength oscillations. As time passes, the short wavelength oscillations cover all the space profile of the wave and, as a result, if the energy per particle is large enough, thermalization is observed. Such a phenomenonology resembles the one of tygers in the Galerkin-truncated Burgers equation \cite{Majda2000,Ray2011,Ray2015} with a crucial difference: the tygers appear far away from the spatial point of the shock, while in our case the short wavelength oscillations are originated at the shock point. However, the route to thermalization after the shock time is qualitatively similar and possible analogies deserve further investigations. 

Our approach extends well beyond the FPUT problem. Adapting the arguments of \cite{Ray2011}, it yields the formation time of tygers in the Galerkin-truncated Burgers equation when a large number of Fourier modes is initially excited (see \cite{supplemental}). This probabilistic approach naturally generalizes to the study of extremal statistics in nonlinear systems, both at thermal equilibrium and  far-from-equilibrium, as in the case considered extensively in this Letter.

Thermalization in the FPUT chain remains an open and debated problem. Several approaches have been proposed to describe its long-time dynamics. For certain classes of initial data, heuristic arguments based on wave turbulence theory predict specific thermalization time scales \cite{Onorato2015}, whereas other initial conditions generate periodic solutions, known as $q$-breathers \cite{Flach2005,Christodoulidi2010}, which never thermalize. The picture becomes even subtler in near-integrable regimes, where the network of interactions among actions and phases has been characterized \cite{Danieli2019}.

The probabilistic framework presented here offers a novel perspective on this long-standing problem by leveraging recent advancements in the statistical characterization of random functions. Indeed, these methodologies are applicable to a broad spectrum of nonlinear problems, provided a functional mapping can be established between the initial probability distribution and its time-evolved counterpart.

\begin{acknowledgments}
This work was funded by the European Union - Next Generation EU. Views and opinions expressed are however those of the authors only and do not necessarily reflect those of the European Union or the European Commission. Neither the European Union nor the European Commission can be held responsible for them. M.G., R.G. and A.P.~acknowledge support by INdAM (GNFM and GNAMPA).  R.G. acknowledges financial support from the GNAMPA project ``Deterministic and probabilistic evolution of out-of-equilibrium Hamiltonian systems''. S.R.~acknowledges support from the MUR PRIN2022 project BECQuMB Grant No.~20222BHC9Z. E.D.~acknowledges ENS-Lyon for financial supporting of his visit to SISSA and SISSA for the hospitality.
\end{acknowledgments}

%

\newpage

\begin{widetext}

\newpage

\centerline{ \huge SUPPLEMENTAL MATERIAL}

\vspace{0.2cm}
\centerline{ \large to}

\vspace{0.2cm}
\centerline{ \Large Random initial data and average shock time} \centerline{\Large in the
Fermi-Pasta-Ulam-Tsingou chain}

\vspace{0.2cm}
\centerline{ \large Matteo Gallone, Ricardo Grande, Antonio Ponno, Stefano Ruffo and Erwan Druais}

\vspace{0.5cm}
\noindent
In this Supplemental Material we extend the derivation of the shock-time asymptotics for random phases presented in the main text. We provide detailed proofs of the upper and lower bounds, including all intermediate steps. All quantities appearing in the main text are rigorously defined here, with their role in the derivation made explicit. To improve readability, certain rescalings are introduced to avoid cumbersome formulas involving constants and parameters.
For clarity, we adopt the following notation: $\mathbb{E}$ denotes expectation, $\mathbb{T}$ the interval $[0,1]$ with periodic boundary conditions ($0\sim 1$), and the superscript $\omega$ indicates the realization of the random process. In particular, $\lambda_0^\omega$ coincides with the function $\lambda(x,0)$ of the main text.

We then proceed to discuss the difference between the route to thermalization in the FPUT chain and the phenomenon of ``tygers'' in the Galerkin-truncated Burgers equation. In particular, we provide a sketch of the derivation of the formation times of``tygers'', as well as their average, for initial data with many random Fourier modes.

We conclude with a short section discussing the numerical analysis of the shock time.

\section{Asymptotics of the shock time for random initial data}
\renewcommand{\theequation}{S.\arabic{equation}}
\setcounter{equation}{0}

The normalized initial data, up to a reminder of order $\sqrt{\epsilon}$, are
\begin{eqnarray}
	\lambda_{0,p}^\omega(x)&=&\sum_{k=1}^p \frac{1}{\sqrt{p}} \big(\sin(\psi_k^\omega) \cos(2 \pi k x + \theta_k^\omega)+\cos(\psi_k^\omega) \sin(2 \pi k x + \phi_k^\omega) \big) \, , \\
	\rho_{0,p}^\omega(x)&=& \sum_{k=1}^p \frac{1}{\sqrt{p}} \big(\sin(\psi_k^\omega) \cos(2 \pi k x + \theta_k^\omega)-\cos(\psi_k^\omega) \sin(2 \pi k x + \phi_k^\omega) \big) \, .
\end{eqnarray}
Given
\begin{equation}
	(\tau_{s,\lambda,p}^\omega)^{-1} := \max_{x \in \mathbb{T}} \partial_x \lambda_{0,p}^\omega(x) \, , \qquad (\tau_{s,\rho,p}^\omega)^{-1} := \max_{x \in \mathbb{T}} \partial_x \rho_{0,p}^\omega(x) \, ,
\end{equation}
we define the shock time of the system as
\begin{equation}\label{eq:DefTsApp}
	\tau_{s,p}^\omega := \min\{\tau_{s,\lambda,p}^\omega, \tau_{s,\rho,p}^\omega\} \, .
\end{equation}
As a consequence of the fact that $\tau_{s,\lambda,p}^\omega, \tau_{s,\rho,p}^\omega\geq 0$, 
\begin{equation}
	(\tau_{s,p}^\omega)^{-1} = \max \{ (\tau_{s,\lambda,p}^\omega)^{-1}, (\tau_{s,\rho,p}^\omega)^{-1} \} \leq (\tau_{s,\lambda,p}^{\omega})^{-1}+(\tau_{s,\rho,p}^\omega)^{-1} \, .
\end{equation}
Since $\tau_{s,\lambda,p}^\omega$ and $\tau_{s,\rho,p}^\omega$ have the same probability distribution, we have
\begin{equation}\label{eq:TsAndTsLambda}
	\mathbb{E} \left[(\tau_{s,p}^\omega)^{-1} \right] \leq 2 \mathbb{E} \left[(\tau_{s,\lambda,p}^\omega)^{-1} \right] \, .
\end{equation}
To simplify the analysis, we introduce the following random variables
\begin{eqnarray}
	X_p^\omega &:=& \max_{x \in \mathbb{T}} \sum_{k=1}^p \frac{k}{\sqrt{p}} \sin(\psi_k^\omega) \sin(2 \pi k x + \theta_k^\omega) \, , \\
	Y_p^\omega &:=& \max_{x \in \mathbb{T}} \sum_{k=1}^p \frac{k}{\sqrt{p}} \cos(\psi_k^\omega) \cos(2 \pi k x + \phi_k^\omega) \, .
\end{eqnarray}
which are related to the $\lambda$ and $\rho$ fields as follows:
\begin{eqnarray}
	X_p^\omega &=& \frac{1}{2} \max_{x \in \mathbb{T}} \big(\partial_x \lambda_{0,p}^\omega(x)+ \partial_x \rho_{0,p}^\omega(x) \big) \, , \\
	Y_p^\omega&=& \frac{1}{2} \max_{x \in \mathbb{T}} \big(\partial_x \lambda_{0,p}^\omega(x)-\partial_x \rho_{0,p}^\omega(x) \big) \, .
\end{eqnarray}
The expectation of the shock time and $\mathbb{E}[X_p^\omega]$ are proportional, as shown by the following inequalities:
\begin{equation}\label{eq:IneqXTs}
	\mathbb{E}[X_p^\omega] \leq \mathbb{E}[ (\tau_{s,p}^\omega)^{-1}] \leq 4 \mathbb{E}[X_p^\omega] \, .
\end{equation}
Indeed, 
\begin{equation}
	\begin{split}
		\mathbb{E}[X_p^\omega] &= \frac{1}{2} \mathbb{E}\Big[\max_{x \in \mathbb{T}}( \partial_x \lambda_{0,p}^\omega(x)-\partial_x \rho_{0,p}^\omega(x))\Big] \leq \frac{1}{2} \mathbb{E}\big[(\tau_{s,\lambda,p}^\omega)^{-1}+(\tau_{s,\rho,p}^\omega)^{-1}\big] \\
		& \overset{\text{\eqref{eq:DefTsApp}}}{\leq} \mathbb{E}\big[(\tau_{s,p}^\omega)^{-1}\big] \overset{\text{\eqref{eq:TsAndTsLambda}}}{\leq} 2 \mathbb{E}[(\tau_{s,\lambda,p}^\omega)^{-1}] \leq 2 \mathbb{E}[X_p^\omega+Y_p^\omega] \leq 4 \mathbb{E}[X_p^\omega] \, ,
	\end{split}
\end{equation}
where in the last step we used the fact that $X_p^\omega$ and $Y_p^\omega$ have the same probability distribution. As a consequence of \eqref{eq:IneqXTs}, we know that to estimate $\mathbb{E}[(\tau_{s,p}^\omega)^{-1}]$ it will be sufficient to estimate $\mathbb{E}[X_p^\omega]$.

To estimate $\mathbb{E}[X_p^\omega]$, we introduce a new random variable
\begin{equation}
	\tilde{\theta}_k^\omega := \left\{\begin{array}{lcl}
		\theta_k^\omega + \pi & \qquad & \text{if $\theta_k^\omega < \pi$} \\
		\theta_k^\omega - \pi & & \text{if $\theta_k^\omega \geq \pi$}
	\end{array}\right.
\end{equation}
which is uniformly distributed in $[0,2\pi)$. Then, 
\begin{equation}
	\sin(\psi_k^\omega) \sin(2 \pi k x + \theta_k^\omega) = \sin(\psi_k^\omega) \cos(2 \pi k x + \tilde{\theta}_k^\omega) \, .
\end{equation}
We also define
\begin{equation}
	f_p^\omega(x) := \sum_{k=1}^p k \sin(\psi_k^\omega) \cos(2 \pi k x + \tilde{\theta}_k^\omega) \, ,
\end{equation}
so that $\max_{x \in \mathbb{T}} f_p^\omega(x)$ has the same probability distribution as $\sqrt{p} X_p^\omega$. In particular,
\begin{equation}\label{eq:FandX}
	\mathbb{E}\Big[\max_{x \in \mathbb{T}} f_p^\omega(x) \Big] = \sqrt{p} \, \mathbb{E}[X_p^\omega] \, .
\end{equation}
At this point, we reduced the problem of estimating $\mathbb{E}[(\tau_{s,p}^\omega)^{-1}]$ to the problem of estimating $\mathbb{E}[\max_{x \in \mathbb{T}} f_p^\omega(x) ]$.

\subsection{Upper bound: First canonical distance of the random process $\{f_p^\omega(x)\}$}
In this section, we introduce the first canonical distance of the random process $\{f_p^\omega(x)\}$. The latter is a distance on $\mathbb{T}$ induced by the random process $\{f_p^\omega\}_{x \in \mathbb{T}}$ given by
\begin{equation}\label{eq:upper-canonical-distance}
	d_p(x,y):=\sqrt{\mathbb{E}\Big[\big(f_p^\omega(x)-f_p^\omega(y)\big)^2\Big]} \, , \qquad x,y \in \mathbb{T} \, .
\end{equation}
As we shall show in a while, the geometric properties of the space $(\mathbb{T},d_p)$ are related to the statistics of $\mathbb{E}[\max_{x \in \mathbb{T}} f_p^\omega(x)]$. The key ingredient is the \emph{covering number} $N_p(\varepsilon)$, which is the smallest number of balls in the (pseudo-)distance $d_p$ of radius $\varepsilon>0$ that cover $\mathbb{T}$.

If $\{f_p^\omega(x)\}_{x \in \mathbb{T}}$ were a Gaussian process,
we could rely on Dudley's result \cite{Dudley1967,Talagrand2014} to have
\begin{equation}\label{eq:DudleyDodo}
	\mathbb{E}\Big[\max_{x \in \mathbb{T}} f_p^\omega(x)\Big] \lesssim \int_0^{\mathrm{diam}_p(\mathbb{T})/2} \sqrt{\log N_p(\varepsilon)} \, d \varepsilon \, .
\end{equation}
In our case, such a bound does not apply straightforwardly because our process is not Gaussian. We will now show that a bound like \eqref{eq:DudleyDodo} holds for our process despite the lack of Gaussianity. 

Firstly, note that we have an explicit expression for $d_p(x,y)$. More precisely,
\begin{equation}\label{eq:Dfexplicit}
	d_p(x,y)= \sqrt{\frac{1}{2} \sum_{k=1}^p k^2\Big(1-\cos\big(2 \pi k (x-y)\big)\Big)} \, .
\end{equation}
Indeed, starting from the definition,
\begin{equation}\label{eq:PrimoPassoDf}
	\begin{split}
		(d_p(x,y))^2&= \mathbb{E}[(f_p^\omega(x)-f_p^\omega(y))^2] \\
		&=\mathbb{E}\big[(f_p^\omega(x))^2+(f_p^\omega(y))^2 - 2 f_p^\omega(x) f_p^\omega(y)\big] \\
		&=2 \mathbb{E}\big[(f_p^\omega(x))^2\big] - 2 \mathbb{E}\big[f_p^\omega(x) f_p^\omega(y)\big] \, ,
	\end{split}
\end{equation}
where in the last step we used the fact that the process $f_p^\omega$ is stationary. Denoting now with $S_a(p)$ the quantity
\begin{equation}
	S_a(p)=\sum_{k=1}^p k^a \overset{p \to +\infty}{\sim} p^{a+1} \, ,
\end{equation}
we have
\begin{equation}\label{eq:Fpomegaquadrato}
	\mathbb{E}[(f_p^\omega(x))^2] = \sum_{k=1}^p k^2 \mathbb{E}[\sin^2(\psi_k^\omega)] \mathbb{E}[\cos^2(2 \pi k x + \theta_k^\omega)] = \frac{1}{4} \sum_{k=1}^p k^2 =\frac{1}{4} S_2(p) \, ,
\end{equation}
\begin{equation}\label{eq:Fpomegaxy}
	\begin{split}
		\mathbb{E}[f_p^\omega(x) f_p^\omega(y)] &= \sum_{k=1}^p k^2 \mathbb{E}[\sin^2 (\psi_k^\omega)] \mathbb{E}[\cos(2 \pi k x+\theta_k^\omega) \cos(2 \pi k y + \theta_k^\omega)]  \\
		&=\sum_{k=1}^p k^2 \frac{1}{2} \mathbb{E}\left[\frac{\cos(2 \pi k (x+y)+2 \theta_k^\omega)}{2}+\frac{\cos(2 \pi k (x-y))}{2} \right] \\
		&=\frac{1}{4} \sum_{k=1}^p k^2 \mathbb{E}[\cos(2 \pi k (x-y))] =\frac{1}{4} \sum_{k=1}^p k^2 \cos(2 \pi k (x-y)) \, .
	\end{split}
\end{equation}
Inserting \eqref{eq:Fpomegaquadrato} and \eqref{eq:Fpomegaxy} into \eqref{eq:PrimoPassoDf}, we obtain
\begin{equation}\label{eq:Upper-CiuchinoMangiaGrano}
	d_p(x,y)^2 = \frac{1}{2} S_2(p) - \frac{1}{2} \sum_{k=1}^p k^2 \cos(2 \pi k (x-y)) = \frac{1}{2} \sum_{k=1}^p k^2(1-\cos(2 \pi k(x-y))) \, ,
\end{equation}
which yields \eqref{eq:Dfexplicit}.

As a next step, we compare the distance $d_p$ with the Euclidean distance $d_e$ defined as
\begin{equation}
	d_e(x,y) := \min_{m \in \mathbb{Z}}|x-y+m| \, .
\end{equation}
Using that, for any $\delta \in (0,2)$ we have $1-\cos(2 \pi k x) \leq 4 \pi^2 (k x)^\delta$, we get
\begin{equation}\label{eq:DfDeEsteso}
	(d_p(x,y))^2 \lesssim \sum_{k=1}^p k^2 (k d_e(x,y))^\delta = (d_e(x,y))^\delta \sum_{k=1}^p k^{2+\delta} = S_{2+\delta}(p) (d_e(x,y))^\delta \, .
\end{equation}
The connection between $d_p$ and $d_e$ will be useful in order to estimate the metric entropy. To do so, let us denote by $M_\eta (\varepsilon)$ the packing number with respect to the distance $d_\eta$, $\eta=e,p$. The packing number is the greatest possible number of elements in an $\varepsilon$-distinguishable subset of $\mathbb{T}$. Recall that an $\varepsilon$-distinguishable set $\{x_j\}_{j \in J}$ is a set of elements $x_j \in \mathbb{T}$ such that $d_\eta(x_j,x_{j'})\geq \varepsilon$ for all $j \neq j'$ in $J$.

Note that, by definition, $M_\eta(\varepsilon) \leq N_\eta(\varepsilon) \leq M_\eta(\varepsilon/2)$. By \eqref{eq:DfDeEsteso}, we have that for any $\delta\in (0,2)$
\begin{equation}
	d_p(x,y) \lesssim d_e(x,y)^{\frac{\delta}{2}} \sqrt{S_{2+\delta}(p)},
\end{equation}
and therefore every $\varepsilon$-separated set in $(\mathbb{T},d_p)$ is $\varepsilon^{\frac{2}{\delta}} (S_{2+\delta}(p))^{-\frac{1}{\delta}}$-separated in $(\mathbb{T},d_e)$. In other words,
\begin{equation}
	M_p(\varepsilon) \leq M_e \left(\left( \frac{\varepsilon}{\sqrt{S_{2+\delta}(p)}}\right)^{\frac{2}{\delta}} \right) \lesssim \left(\frac{\varepsilon}{\sqrt{S_{2+\delta}(p)}} \right)^{-\frac{2}{\delta}} \, .
\end{equation}
As a neat result,
\begin{equation}\label{eq:EntropyEstimateeee}
	N_p(\varepsilon) \lesssim \left(\frac{\sqrt{S_{2+\delta}(p)}}{\varepsilon} \right)^{\frac{2}{\delta}} \, , \qquad \forall \varepsilon>0,\  0<\delta<2 \, .
\end{equation}

\subsection{Upper bound: chaining argument and Dudley's bound}
As we already mentioned, to bound from above $\mathbb{E}[\max_{x \in \mathbb{T}} f_p^\omega(x)]$ we can not use directly Dudley's bound because our random variables are not Gaussian. The key idea to prove a bound like \eqref{eq:DudleyDodo} exploits the intuition that the tails of the probability distribution of $\max_{x \in \mathbb{T}} f_p^\omega(x)$ are subGaussian. This is to say that, in the end, the estimate will rely on an estimate of the tails
\begin{equation}
	\mathbb{P}\Big(\max_{x \in \mathbb{T}} f_p^\omega(x) > \lambda\Big) \, , \qquad \text{as } \lambda \to +\infty \, .
\end{equation}

Let us recall that, if $Z^\omega$ is a nonnegative random variable, by Cavalieri's principle
\begin{equation}
	\mathbb{E}[Z^\omega] = \int_0^{+\infty} \mathbb{P}(Z^\omega > \lambda) \, d\lambda \, .
\end{equation}
Thinking of the integral as the limit of a Riemann sum, we might start by finding a good bound for $\mathbb{P}(\max_{i=1,\dots,n} f_p^\omega(x_i) > \lambda)$. Using a simple ``union bound'', one has
\begin{equation}\label{eq:ToBeImproved}
	\mathbb{P}\Big(\max_{i=1,\dots,n} f_p^\omega(x_i) > \lambda\Big) \leq \sum_{i=1}^n \mathbb{P}(f_p^{\omega}(x_i) > \lambda) \, .
\end{equation}
This latter bound, as it is, is not precise enough because the RHS diverges as $n \to +\infty$. In order to make the idea work and to consider the bound over an infinite set, one needs to exploit two ingredients:
\begin{itemize}
	\item[(i)] continuity of the function $x \mapsto f_p^\omega(x)$
	\item[(ii)] a large overlap between some sets $\{f_p^\omega(x_i)>\lambda\}$ and $\{f_p^\omega(x_j)>\lambda\}$ if $x_i$ and $x_j$ are close. This latter point means that \eqref{eq:ToBeImproved} is wasteful and one can have better bounds exploiting this overlap.
\end{itemize}

We now show that the vicinity of two points in item (ii) is best measured in terms of the first canonical distance. In particular, we prove that
\begin{equation}\label{eq:TesiLemma1}
	\mathbb{P}\big(|f_p^\omega(x)-f_p^\omega(y)|>\lambda\big) \leq 2 \exp \left(-\frac{\lambda^2}{8 d_p(x,y)^2} \right) \, .
\end{equation}
Indeed, we have
\begin{equation}\label{eq:EquationF}
	\begin{split}
		f_p^\omega(x)-f_p^\omega(y) &=\sum_{k=1}^p k \sin(\psi_k^\omega) \Big(\cos(2 \pi k x + \theta_k^\omega)-\cos(2 \pi k y+\theta_k^\omega) \Big) \\
		&=\sum_{k=1}^p 2 k \sin(\psi_k^\omega) \sin\Big({\textstyle \frac{2 \pi k (y-x)}{2}}\Big) \sin\Big({\textstyle \frac{2 \pi k (x+y)}{2}+\theta_k^\omega}\Big) \, .
	\end{split}
\end{equation}
For any $z>0$, we have the following Chernoff bound 
\begin{equation}\label{eq:Chernoff}
	\begin{split}
		\mathbb{P}(f_p^\omega(x)-f_p^\omega(y) > \lambda) &\leq e^{-\lambda z} \mathbb{E} \left[ \exp \left(\sum_{k=1}^p 2 k z \sin (\psi_k^\omega) \sin\Big({\textstyle \frac{2 \pi k(y-x)}{2}}\Big)\sin\Big(\theta_k^\omega+{\textstyle \frac{2 \pi k (x+y)}{2}}\Big) \right) \right] \\
		&= \exp(-\lambda z) \prod_{k=1}^p \mathbb{E}\left[ \exp \left( 2 k z \sin (\psi_k^\omega) \sin\big({\textstyle \frac{2 \pi k(y-x)}{2}}\big)\sin\big(\theta_k^\omega+{\textstyle \frac{2 \pi k (x+y)}{2}}\big)\right)\right]
	\end{split}
\end{equation}
Let us note that for any $\alpha \in \mathbb{R}$, we have
\begin{equation}
	\begin{split}
		\mathbb{E}\big[\exp(\alpha \sin(\psi_k^\omega) \sin(\theta_k^\omega))\big]&= \frac{1}{4 \pi^2} \int_0^{2 \pi} \int_0^{2 \pi} e^{\alpha \sin \theta \sin \psi} \, d\psi d \theta \\
		&= \frac{1}{2 \pi^2} \int_0^\pi \int_0^\pi (e^{\alpha \sin \theta \sin \psi}+e^{-\alpha \sin \theta \sin \psi}) \, d\theta d \psi \\
		&=\frac{1}{\pi^2} \int_0^\pi \int_0^\pi \cosh(\alpha \sin \theta \sin \psi) \, d\theta d \psi \\
		&\leq \frac{1}{\pi^2}\int_0^\pi \int_0^\pi \exp\left(\frac{\alpha^2}{2}\, \sin^2\psi\, \sin^2 \theta\right) \, d \theta d \psi \leq e^{\frac{\alpha^2}{2}}
	\end{split}
\end{equation}
Using the latter inequality in \eqref{eq:Chernoff}, we have
\begin{equation}
	\begin{split}
		\mathbb{P}(f_p^\omega(x)-f_p^\omega(y) > \lambda) & \leq e^{-\lambda z} \prod_{k=1}^p \exp\left(\frac{4 k^2 z^2 \sin^2(\frac{2 \pi k (y-x)}{2})}{2} \right) \\
		&\leq \exp\left(-\lambda z +z^2 \sum_{k=1}^p 2 k^2 \sin^2\big({\textstyle \frac{2 \pi k (y-x)}{2}}\big) \right) \\
		&=\exp\left(-\lambda z+z^2 \sum_{k=1}^p(1-\cos(2 \pi k (y-x))) \right) = \exp(-\lambda z + 2 z^2 d_p(x,y)^2)
	\end{split}
\end{equation}
Choosing now $z=\frac{\lambda}{4 d_p(x,y)^2}$ yields \eqref{eq:TesiLemma1}.

At this point, we are ready to give the main strategy to bound $\mathbb{P}(\max_{x \in \mathbb{T}} f_p^\omega(x))$. The idea is to construct finite subsets $\mathcal{N}_n \subset \mathbb{T}$ whose points are closer as closer when $n$ increases. For $r \in (0,1)$, choose $\mathcal{N}_n$ such that 
\begin{equation}
	d_p(x,\mathcal{N}_n) \lesssim r^n \, ,\qquad \forall x \in \mathbb{T} \, , \qquad \forall n \in \mathbb{N} \, .
\end{equation}
The existence of the set $\mathcal{N}_n$ with such properties is guaranteed by the compactness of $(\mathbb{T},d_p)$.

On $\mathbb{T}$, we define projections $\pi_n:\mathbb{T} \to \mathcal{N}_n$ so that $\pi_n(x)$ is the closest point to $x$ in $\mathcal{N}_n$, with distance measured in $d_p$. With these tools at hand, we can decompose $f_p^\omega$ along the chain $\{\pi_n(x)\}_{n\in\mathbb{N}_0}$, i.e.
\begin{equation}
	f_p^\omega(x) = f_p^\omega(\pi_0(x))+\sum_{n=1}^{\infty} \big(f_p^\omega(\pi_n(x))-f_p^\omega(\pi_{n-1}(x))\big) \, .
\end{equation}
Then,
\begin{equation}
	\max_{x \in \mathbb{T}} f_p^\omega(x) \leq \max_{x \in \mathbb{T}} f_p^\omega(\pi_0(x))+\sum_{n=1}^\infty \max_{x \in \mathbb{T}} \left(f^\omega(\pi_n(x))-f^\omega(\pi_{n-1}(x)) \right)
\end{equation}
where now each supremum is over a maximum over a finite set.

To estimate each of the differences $f_p^\omega(\pi_n(x))-f_p^\omega(\pi_{n-1}(x))$ we will use \eqref{eq:TesiLemma1}. For the sum $\sum_{n=1}^\infty$ we will use the following idea. If $Z_n^\omega$ is a random variable,
\begin{equation}
	\mathbb{P}\Big(\sum_{n=1}^\infty |Z_n^\omega| > \lambda\sigma\Big) \leq \sum_{n=1}^\infty \mathbb{P}\big(|Z_n^\omega| > \lambda \sigma_n\big)
\end{equation}
where $(\sigma_n)_{n=1}^\infty \subset \mathbb{R}_+$ is any sequence such that $\sigma=\sum_{n=1}^\infty \sigma_n <\infty$. For $r\in (0,\frac18)$ sufficiently small, set
\begin{equation}
	\sigma_n = (8r)^n \sqrt{\log(2^n |\mathcal{N}_n| |\mathcal{N}_{n-1}|)} \, , \qquad n \in \mathbb{N} \, .
\end{equation}
Since $\sigma=\sum_{n=1}^\infty \sigma_n < +\infty$, which we explain in \eqref{eq:sigma_finite} below, we have
\begin{equation}
	\begin{split}
		\mathbb{P}\big(\max_{x \in \mathbb{T}} |f_p^\omega(x)-f_p^\omega(0)|> \lambda \sigma\big) & \leq \mathbb{P}\Big(\sum_{n=1}^\infty \max_{x \in \mathbb{T}} |f_p^\omega(\pi_n(x))-f_p^\omega(\pi_{n-1}(x))> \lambda \sigma\Big) \\
		&\leq \sum_{n=1}^\infty \mathbb{P}\big(\exists x \in \mathbb{T} \, : \, |f_p^\omega(\pi_n(x))-f_p^\omega(\pi_{n-1}(x))| > \lambda \sigma_n\big) \\
	&\lesssim \sum_{n=1}^\infty |\mathcal{N}_n| | \mathcal{N}_{n-1}| \exp\left(-\frac{(\lambda \sigma_n)^2}{8 (d_p(\pi_n(x), \pi_{n-1}(x))^2} \right)\\
		&\leq \sum_{n=1}^\infty |\mathcal{N}_n| |\mathcal{N}_{n-1}| (|\mathcal{N}_{n}| | \mathcal{N}_{n-1}|)^{-\lambda^2} 2^{-n \lambda^2} \\
	&\lesssim \sum_{n=1}^\infty 2^{-n \lambda^2} \lesssim 2^{-\lambda^2} \, , \qquad \text{for $\lambda>1$} \, .
	\end{split}
\end{equation}
Finally, we obtain
\begin{equation}
	\begin{split}
		\mathbb{E}\big[\max_{x \in \mathbb{T}} |f_p^\omega(x)-f_p^\omega(0)|\big] &=\int_0^\infty \mathbb{P}\Big(\max_{x \in \mathbb{T}} |f_p^\omega(x)-f_p^\omega(0)| > \lambda\Big) \, d\lambda \\
		&=\sigma \int_0^\infty \mathbb{P}\Big(\max_{x \in \mathbb{T}} |f_p^\omega(x)-f_p^\omega(0)| > \sigma \tilde{\lambda}\Big) \, d \tilde{\lambda} \\
		&\lesssim \sigma \left( \int_0^1 d\tilde{\lambda}+\int_1^\infty 2^{-\tilde{\lambda}^2} \, d \tilde{\lambda}\right) \lesssim \sigma \, .
	\end{split}
\end{equation}
Similarly,
\begin{equation}
	\mathbb{E}\Big[\max_{x \in \mathbb{T}} f_p^\omega(x)\Big] \leq \mathbb{E}\Big[\max_{x \in \mathbb{T}} \big|f_p^\omega(x)-f_p^\omega(0)\big|\Big]+\mathbb{E}\big[|f_p^\omega(0)|\big] \lesssim \sigma + \sqrt{S_2(p)} \, .
\end{equation}
Since $\sqrt{S_2(p)} \sim p^{\frac{3}{2}}$, we need to estimate only $\sigma$. Indeed,
\begin{equation}\label{eq:sigma_finite}
	\begin{split}
		\sigma &\lesssim \sum_{n=1}^\infty (8r)^n \sqrt{\log (2^n |\mathcal{N}_n ||\mathcal{N}_{n-1}|)} \\
		&\leq \sum_{n=1}^\infty (8r)^n (\sqrt{n} \sqrt{\log 2} + \sqrt{\log|\mathcal{N}_n|} + \sqrt{\log|\mathcal{N}_{n-1}|}) \\
		&\lesssim \sum_{n=1}^\infty (8r)^n \sqrt{\log| \mathcal{N}_n|} \sim \int_0^{\mathrm{diam}_p(\mathbb{T})/2} \sqrt{\log N_p(\varepsilon)} \, d\varepsilon
	\end{split}
\end{equation}
and we are left in estimating this latter integral. Indeed, noting that $\mathrm{diam}_p(\mathbb{T})/2 \sim \sqrt{S_2(p)}$, we get. 
\begin{equation}
	\begin{split}
		\sigma & \lesssim \int_0^{\sqrt{S_2(p)}} \sqrt{\log N_p(\varepsilon)} \, d\varepsilon \\
		&\hspace{-0.2cm}\overset{\text{\eqref{eq:EntropyEstimateeee}}}{\lesssim} \int_0^{\sqrt{S_2(p)}} \sqrt{\frac{2}{\delta} \log \left(\frac{\sqrt{S_{2+\delta}(p)}}{\varepsilon} \right)} \, d \varepsilon \\
		&\hspace{-0.8cm}\overset{\varepsilon=\sqrt{S_{2+\delta}(p)}e^{-z}}{=} \frac{2}{\sqrt{\delta}} \int_{\log \sqrt{\frac{S_{2+\delta}(p)}{S_2(p)}}}^{+\infty} \sqrt{z} e^{-z} \sqrt{S_{2+\delta}(p)} \, d z \\
		&\lesssim \frac{1}{\sqrt{\delta}} \sqrt{S_{2+\delta}(p)} \int_0^{+\infty} \sqrt{z} e^{-z} \, dz \\
		&\lesssim \frac{1}{\sqrt{\delta}} \sqrt{S_{2+\delta}(p)} \\
		&\lesssim \frac{1}{\sqrt{\delta}} p^{\frac{3}{2}+\frac{\delta}{2}}  \end{split}
\end{equation}
Choosing now $\delta=2 (\log p)^{-1}$, we have $p^\delta=p^{(\log p)^{-1}}=e$. This proves the following
\begin{equation}
	\mathbb{E}\Big[\max_{x \in \mathbb{T}} |f_p^\omega(x)|\Big] \lesssim p^{\frac{3}{2}} \sqrt{\log p} \, .
\end{equation}
Whence, using \eqref{eq:FandX}, we get $\mathbb{E}[X_p^\omega] \lesssim p \sqrt{\log p}$ and, by \eqref{eq:IneqXTs} we complete the proof of the upper bound
\begin{equation}
	\mathbb{E}\big[(\tau_{s,p}^\omega)^{-1}\big] \lesssim p \sqrt{\log p} \, .
\end{equation}

\subsection{Lower bound: second canonical distance}
The construction of a sharp lower bound for $\mathbb{E}[\max_{x \in \mathbb{T}} f_p^\omega(x)]$ is more delicate than the upper bound. At first we note that
\begin{equation}\label{eq:pingiuno_nucleare}
	\max_{x \in \mathbb{T}} |f_p^\omega(x)| \leq \max_{x \in \mathbb{T}} f_p^\omega(x)+\max_{x \in \mathbb{T}} \big( - f_p^\omega(x)\big) \, .
\end{equation}
Then, since $f_p^\omega(x)$ and $-f_p^\omega(x)$ have the same probability distribution, we can take the expectation on both sides of \eqref{eq:pingiuno_nucleare} to get
\begin{equation}
	\mathbb{E}\Big[ \max_{x \in \mathbb{T}} f_p^\omega(x) \Big] \leq \mathbb{E} \Big[ \max_{x \in \mathbb{T}} f_p^\omega(x) \Big] + \mathbb{E}\Big[ \max_{x \in \mathbb{T}} \big(-f_p^\omega(x)\big) \Big] = 2 \mathbb{E} \Big[ \max_{x \in \mathbb{T}} f_p^\omega(x) \Big] \, ,
\end{equation}
so that
\begin{equation}\label{eq:lower-star}
	\mathbb{E}\Big[ \max_{x \in \mathbb{T}} f_p^\omega(x) \Big] \geq \frac{1}{2} \mathbb{E} \Big[ \max_{x \in \mathbb{T}} |f_p^\omega(x)| \Big] \, .
\end{equation}
We now introduce an auxiliary random process $g_p^\omega$ and conveniently rewrite $f_p^\omega$ as
\begin{equation}\label{eq:lower-asterisco}
	\begin{split}
		f_p^\omega(x)&= \Ree \Big(\sum_{k=1}^p k \sin(\psi_k^\omega ) e^{i \tilde{\theta}_k^\omega} e^{2 \pi i k x} \Big) \, , \\
		g_p^\omega(x)&= \Imm \Big(\sum_{k=1}^p k \sin(\psi_k^\omega) e^{i \tilde{\theta}_k^\omega} e^{2 \pi i j x} \Big) \, .
	\end{split}
\end{equation}
From the explicit expressions of $f_p^\omega$ and $g_p^\omega$, one sees that the two processes have the same probability distribution. Moreover,
\begin{equation}\label{eq:lower-anonima}
	\max_{x \in \mathbb{T}} \sqrt{\big( f_p^\omega(x) \big)^2+\big(g_p^\omega(x) \big)^2} \leq \sqrt{2} \max_{x \in \mathbb{T}} \max\Big\{ |f_p^\omega(x) | , |g_p^\omega(x)| \Big\} \leq \sqrt{2} \Big( \max_{x \in \mathbb{T}} |f_p^\omega(x) | + \max_{x \in \mathbb{T}} |g_p^\omega(x)| \Big) \, .
\end{equation}
Using \eqref{eq:lower-asterisco} and the fact that $g_p^\omega$ and $f_p^\omega$ have the same probability distribution, from \eqref{eq:lower-anonima} we get
\begin{equation}\label{eq:lower-quadrato}
	\mathbb{E} \Big[ \max_{x \in \mathbb{T}} \sqrt{\big( f_p^\omega(x) \big)^2+\big(g_p^\omega(x) \big)^2} \Big] \leq 2 \sqrt{2}\, \mathbb{E} \Big[ \max_{x \in \mathbb{T}} |f_p^\omega(x)| \Big] \, .
\end{equation}
Noting that
\begin{equation}\label{eq:lower-triangolo}
	\mathbb{E}\Big[ \max_{x \in \mathbb{T}} \sqrt{\big( f_p^\omega(x) \big)^2+\big(g_p^\omega(x) \big)^2} \Big] = \mathbb{E} \Big[ \max_{x \in \mathbb{T}} \Big| \sum_{k=1}^p k \sin(\psi_k^\omega) e^{i \tilde{\theta}_k^\omega} e^{2 \pi i k x} \Big| \Big] \, ,
\end{equation}
and combining \eqref{eq:lower-star}, \eqref{eq:lower-quadrato} and \eqref{eq:lower-triangolo} we get
\begin{equation}\label{eq:lower-flower}
	\mathbb{E}\Big[ \max_{x \in \mathbb{T}} f_p^\omega(x) \Big] \geq \frac{1}{2} \mathbb{E}\Big[\max_{x \in \mathbb{T}} | f_p^\omega(x) | \Big] \geq \frac{1}{4 \sqrt{2}} \mathbb{E} \Big[\max_{x \in \mathbb{T}} \Big|\sum_{k=1}^p k \sin(\psi_k^\omega) e^{i \tilde{\theta}_k^\omega} e^{2 \pi i k x} \Big| \Big] \, .
\end{equation}
All that remains is to find a lower bound for the RHS of \eqref{eq:lower-flower}. Since $\tilde{\theta}_k^\omega$ and $\psi_{k'}^\omega$ are independent random variables for every $k$ and $k'$, we can decompose the probability space $\Omega$ as $\Omega=\Omega_1 \times \Omega_2$ in such a way that $\psi_k^\omega=\psi_k^{\omega_1}$ and $\tilde{\theta}_k^\omega=\tilde{\theta}_k^{\omega_2}$. According to this decomposition, the probability measure is factorized as $\mathbb{P}=\mathbb{P}_1 \otimes \mathbb{P}_2$.

Denoting now with $\mathbb{E}_1$ the average over $\Omega_1$ only, we study 
\begin{equation}\label{eq:lower-A}
	\mathbb{E}_1 \Big[ \max_{x \in \mathbb{T}} \Big| \sum_{k=1}^p k \sin(\psi_k^{\omega_1}) e^{i \tilde{\theta}_k^{\omega_2}} e^{2 \pi i k x} \Big| \Big].
\end{equation}
One can define a canonical distance for this process, different from the one defined in \eqref{eq:upper-canonical-distance} (see \cite[Proposition 3.2.15]{Talagrand2014}):
\begin{equation}\label{eq:lower-UUUIO}
	\begin{split}
		\mathsf{d}_p(x,y)^2&:= \sum_{k=1}^p |k e^{i \tilde{\theta}_k^{\omega_2}}|^2 \big( \mathbb{E}_1 \big[|\sin(\psi_k^{\omega_1})| \big] \big)^2 |e^{2 \pi i k x}-e^{2 \pi i k y}|^2 \\
		&=\sum_{k=1}^p k^2 \left( \frac{2}{\pi} \right)^2 2 \left(1-\cos(2 \pi k (x-y)) \right) \, .
	\end{split}
\end{equation}
Note that $\mathsf{d}_p$, in principle, should be a random distance since we only took the average over $\Omega_1$ in \eqref{eq:lower-A}. Remarkably, this latter distance is deterministic (independent of $\omega_2$) and it is also proportional to the distance $d_p$ defined in \eqref{eq:upper-canonical-distance} and induced by $f_p^\omega$.

Using now the fact that $f_p^\omega(x)$ is stationary, i.e.~its distribution law does not depend on $x$, by \cite[Proposition 3.2.15, Theorem 3.1.1 and Lemma 3.1.2]{Talagrand2014}, together with the fact that the distances $\mathsf{d}_p$ and $d_p$ are equal apart from a multiplicative constant independent of $p$, we have
\begin{equation}\label{eq:lower-C}
	\mathbb{E}_1 \Big[ \max_{x \in \mathbb{T}} \Big|\sum_{k=1}^p k \sin(\psi_k^{\omega_1}) e^{i \tilde{\theta}_k^{\omega_2}} e^{2 \pi i k x} \Big| \Big] \gtrsim \int_0^\infty \sqrt{\log N_p(\varepsilon)} \, d \varepsilon \, .
\end{equation}
Since the RHS of \eqref{eq:lower-C} is independent of $\omega_2$, we can take the average over $\Omega_2$ on both sides to get
\begin{equation}\label{eq:lower-D}
	\mathbb{E}\Big[\max_{x \in \mathbb{T}} \Big| \sum_{k=1}^p k \sin(\psi_k^{\omega_1}) e^{i \tilde{\theta}_k^{\omega_2}} e^{2 \pi i k x} \Big| \Big] \gtrsim \int_0^\infty \sqrt{\log N_p(\varepsilon)} \, d \varepsilon \, .
\end{equation}
Thus, inserting \eqref{eq:lower-D} in \eqref{eq:lower-flower} we have proven
\begin{equation}\label{eq:lower-bullone}
	\mathbb{E}\Big[\max_{x \in \mathbb{T}} f_p^\omega(x) \Big] \gtrsim \int_0^\infty \sqrt{\log N_p(\varepsilon)} \, d \varepsilon 
\end{equation}
which is the inverse of the Dudley bound. At this point, to obtain a lower bound for Dudley's integral it is sufficient to estimate the metric entropy from below.

\subsection{Lower bound: bound from below of the metric entropy}
Since we need to trace the dependence on $p$ for large $p$, we can ignore multiplicative constants and consider the distance
\begin{equation}
	d_p(x,y)^2:=\sum_{k=1}^p k^2 \sin^2(\pi k (x-y)) \, ,
\end{equation}
which is the same as \eqref{eq:Upper-CiuchinoMangiaGrano}. By setting $\tilde{f}_p^\omega(x) := p^{-\frac{3}{2}} f_p^\omega(x)$, we obtain the normalized distance induced by $\tilde{f}_p^\omega$ using \eqref{eq:upper-canonical-distance}, that is
\begin{equation}\label{eq:tilde_d}
	\tilde{d}_p(x,y)^2=\frac{1}{p^3} \sum_{k=1}^p k^2 \sin^2(\pi k (x-y)) \, .
\end{equation}
It follows directly from this normalization that
\begin{equation}\label{eq:lower-paraculo}
	\mathbb{E} \Big[ \max_{x \in \mathbb{T}} f_p^\omega(x) \Big] = p^{\frac{3}{2}} \mathbb{E} \Big[ \max_{x \in \mathbb{T}} \tilde{f}_p^\omega(x) \Big] \overset{\text{\eqref{eq:lower-bullone}}}{\gtrsim} p^{\frac{3}{2}} \int_0^{+\infty} \sqrt{\log \tilde{N}_{p}(\varepsilon)} \, d \varepsilon \, ,
\end{equation}
where $\tilde{N}_{p}$ is the metric entropy associated to the distance $\tilde{d}_p$ in \eqref{eq:tilde_d}.
It is clear that if $p|x-y| \leq 1$, then
\begin{equation}
	1 \geq \tilde{d}_p(x,y)^2 \gtrsim p^{-3} \sum_{k=1}^p k^2 ( \pi k |x-y|)^2 \sim p^2 |x-y|^2 \, .
\end{equation}
This estimate allows us to construct a set of points which are $\varepsilon$-separated with respect to $\tilde{d}_p$ by considering
\begin{equation}\label{eq:lower-R}
	x_j = \frac{j}{p} \varepsilon \, , \qquad \text{for } \quad j=1,\dots,\lfloor \varepsilon^{-1} \rfloor \, ,
\end{equation}
(where $\lfloor \cdot \rfloor$ denotes the integer part). Note that, using $p|x-y| \leq 1$, we get
\begin{equation}
	\tilde{d}_p(x_j,x_{j'})^2 \gtrsim p^2 |x_j-x_{j'}|^2 = p^2 \frac{\varepsilon^2}{p^2} |j-j'| \geq \varepsilon^2
\end{equation}
where in the last step we used that $|j-j'| \geq 1$. The set of points in \eqref{eq:lower-R} are thus $\varepsilon$-separated and, as a result, $\tilde{N}_{p}(\varepsilon) \geq \tilde{M}_{p}(\varepsilon) \geq \lfloor \varepsilon^{-1} \rfloor$.

As we will show below, this lower bound is not optimal and leads to $\int_0^\infty \sqrt{\log \tilde{N}_{p}(\varepsilon)} \, d \varepsilon \gtrsim 1$, which – back to the original $f_p^\omega(x)$ via \eqref{eq:lower-paraculo} – would yield the nonoptimal lower bound $\mathbb{E}[\max_{x \in \mathbb{T}} f_p^\omega(x)] \gtrsim p^{\frac{3}{2}}$.

To improve this latter bound, we have to add more $\varepsilon$-separated points to our set $\{x_j\}_{j=1,\dots,\lfloor \varepsilon^{-1} \rfloor}$. In order to do so, let us start from
\begin{equation}
	\begin{split}
		\tilde{d}_p(x,0)^2 &= \frac{1}{p^3} \sum_{k=1}^p k^2 \sin^2(\pi k x) \\
		&= \frac{1}{p^3} \sum_{k=1}^p k^2 \left( \frac{e^{i \pi k x}-e^{- i \pi k x}}{2 i}\right)^2 \\
		&=-\frac{1}{4p} \sum_{k=1}^p k^2 \big( e^{2 \pi k x}+e^{-2 \pi k x} - 2 \big) \\
		&=\frac{1}{2 p^3} \sum_{k=1}^p k^2 - \frac{1}{2p^3} \Big(\sum_{k=1}^p k^2 \cos(2 \pi k x) \ \Big) \, .
	\end{split}
\end{equation}
We now explicitly sum up the geometric series:
\begin{equation}
	\sum_{k=1}^p e^{2 \pi i k x} = \frac{e^{2 \pi (p+1) x}-e^{2 \pi i x}}{e^{2 \pi i x} - 1} \, ,
\end{equation}
and we define the (translated) Dirichlet kernel
\begin{equation}
	D_p(x):=2 \sum_{k=1}^p \cos(2 \pi k x)=\frac{\sin\big((2p+1)\pi x \big)}{\sin(\pi x)}-1.
\end{equation}
By differentiating two times we get
\begin{equation}
	\sum_{k=1}^p k^2 \cos(2 \pi k x) =-\frac{1}{8 \pi^2} D''_p(x) \, .
\end{equation}
Explicitly, one has
\begin{equation}\label{eq:lower-UV}
	D''_p(x)=\frac{\pi^2}{\sin(\pi x)} \left( \left( \frac{2}{\sin^2(\pi x)}-2-4p(1+p) \right) \sin((1+2p) \pi x) - \frac{2(1+2p) \cos((1+2p)\pi x) \cos(\pi x)}{\sin (\pi x)}\right) \, .
\end{equation}
With this expression at hand, we can compute the Taylor expansion for $p|x| \ll 1$ of $(\tilde{d}_p(x,0))^2$:
\begin{equation}
	(\tilde{d}_p(x,0))^2 = \frac{(6p^4+15p^3+10p^2-1)\pi^2 }{30p^2} x^2 - \frac{(1-7p^2+21p^4+21p^5+6p^6)\pi^4}{126p^2} x^4 + O(p^6x^6) \, .
\end{equation}
In particular, taking only leading terms, we get
\begin{equation}
	(\tilde{d}_p(x,0))^2 \sim \frac{\pi^2}{5} p^2 x^2 \, \qquad x \downarrow 0 \, .
\end{equation}
Note now that for the terms in \eqref{eq:lower-UV}
\begin{equation}
	\begin{split}
		\frac{1}{p^3}|D_p''(x)| &\lesssim \frac{1}{(px)^3}+\frac{1}{p^3 x} + \frac{1}{px} + \frac{1}{(px)^2} \lesssim \frac{1}{(px)^3}+\frac{x^2}{p^3 x^3} + \frac{1}{px} + \frac{1}{(px)^2} < \frac{1}{12}
	\end{split}
\end{equation}
provided $|px|>\mu$ for $\mu$ large enough. 
Indeed, for $p$ large enough, the numerator in \eqref{eq:lower-UV} can be controlled by $p^2$ uniformly (one has to use that $|\sin((2p+1) \pi x)|) \leq 1$), 
Note also that
\begin{equation}
	\lim_{p \to +\infty} \frac{1}{2 p^3} \sum_{k=1}^p k^2 = \frac{1}{6}
\end{equation}
and thus, for large $p$ we have
\begin{equation}\label{eq:lower-UnAutostrada}
	\tilde{d}_p(x,0)^2 \geq \frac{1}{6}-\frac{1}{12} \geq \frac{1}{12} \, \qquad \forall |x| > \frac{\mu}{p} \, .
\end{equation}
To sum up, we have proved that if two points $x,y$ are separated in Euclidean distance as $|x-y| > \frac{\mu}{p}$, then they are separated in the distance $\tilde{d}_p(x,y)$, since $\tilde{d}_p(x,y)^2 > \frac{1}{12}$. With this observation at hand, we can extend the family of points constructed in \eqref{eq:lower-R}. Indeed, let us start by renaming that family with a superscript ``1'':
\begin{equation}
	\{x_j\}_{j=1,\dots,\lfloor \varepsilon^{-1} \rfloor}=\left\{ \frac{j \varepsilon}{p} \right\}_{j=1,\dots,\lfloor \varepsilon^{-1} \rfloor} =: \{x_j^{(1)} \}_{j=1, \dots, \lfloor \varepsilon^{-1} \rfloor} \, .
\end{equation}
This family of points will be referred to as the \emph{first cluster}. Using \eqref{eq:lower-UnAutostrada}, we can construct a second cluster
\begin{equation}
	x_j^{(2)}=\frac{1}{p}+\frac{\mu}{p}+\frac{j \varepsilon}{p} \, ,\qquad j=1,\dots,\lfloor \varepsilon^{-1} \rfloor 
\end{equation}
in such a way that the distance between each point in the cluster 1 and in the cluster 2 is large enough:
\begin{equation}
	\tilde{d}_p(x_j^{(1)},x_{j'}^{(2)}) \geq \frac{1}{\sqrt{12}} \, ,\qquad \text{since} \qquad |x_j^{(1)}-x_{j'}^{(2)}| \geq \frac{\mu}{p} \, , \quad \forall j,j'
\end{equation}
and also, by construction
\begin{equation}
	\tilde{d}_p(x_j^{(2)},x_{j'}^{(2)}) \geq \varepsilon \, ,
\end{equation}
provided $\varepsilon<\varepsilon_{\ast}:=\frac{1}{\sqrt{12}}$.

Repeating the same argument, we can add several clusters and iteratively repeat the construction until we run out of space. In total, we can create $\frac{p}{\mu}$ clusters, each of them with $\varepsilon^{-1}$ points. In total, we have shown that for $p>p_*>\mu$ independent of all the parameters of the system,
\begin{equation}\label{eq:lower-EntropyFromBelow}
	\tilde{N}_{p}(\varepsilon) \geq \frac{p}{\mu} \varepsilon^{-1} \, ,
\end{equation}
provided $\varepsilon < \varepsilon_*=\frac{1}{\sqrt{12}}$.

\subsection{Lower bound: bound from below of Dudley's integral and shock time}
We are now ready to prove the lower bound for Dudley's integral.
\begin{equation}
\label{eq:lower-SantissimoIntegrale}
	\begin{split}
		\int_0^{\varepsilon_*} \sqrt{\log \tilde{N}_{p}(\varepsilon)} \, d \varepsilon &\overset{\text{\eqref{eq:lower-EntropyFromBelow}}}{\gtrsim_\mu} \int_0^{\varepsilon_*} \sqrt{\log(p/\varepsilon)} \, d \varepsilon \\
		&\hspace{-0.2cm}\overset{\varepsilon=p e^{-z}}{=}p \int_{\log (p/\varepsilon_*)}^\infty \sqrt{z} e^{-z} \, dz \\
		&= p \sqrt{\log(p/\varepsilon_*)} e^{-\log(p/\varepsilon_*)} + \frac{1}{2} \sqrt{\pi} \mathrm{Erfc} \, (\sqrt{\log(p/\varepsilon_*)}) \\
		&\gtrsim \sqrt{\log p}
	\end{split}
\end{equation}
At this point, we obtain a lower bound on the shock time by combining \eqref{eq:IneqXTs}, \eqref{eq:FandX}, \eqref{eq:lower-paraculo} and \eqref{eq:lower-SantissimoIntegrale}, we get
\begin{equation}
	\mathbb{E} \Big[ (\tau_{s,p}^{\omega})^{-1} \Big] \gtrsim p \sqrt{\log p} \, .
\end{equation}

\section{Space profile of the wave and differences with tygers}

There is a wide literature on thermalization of the Galerkin truncated Burgers equation, expecially with respect to one of the typical observed phenomenon, called tygers. Given $K>0$, the Galerkin truncated Burgers equation is the equation
\begin{equation}\label{eq:TruncatedBurgers}
    \partial_t u - P_{K}\left(u\partial_x u\right)=0 \, ,
\end{equation}
where $P_K$ is the Fourier projector on the modes $\{-K,-K+1, \cdots,K-1,K\}$. Such an equation has very interesting features that can be summarized as
\begin{itemize}
    \item It is conservative and, in particular, it conserves both \emph{momentum} $M$ and \emph{energy} $E$ defined as
    \begin{eqnarray}
        M&:=&  \int_0^1 u(x) \, dx \, \\
        E&:=&\frac{1}{2} \int_0^1 u^2(x) \, dx \, .
    \end{eqnarray}
    See, for example, \cite{Majda2000}.
    \item The equation has a family of invariant Gibbs measures labeled by a parameter $\beta>0$ which plays the role of temperature (see eq.~9 in \cite{Majda2000}).
    \item As a consequence of Galerkin truncation, short before the shock time there appears rapid oscillations called \emph{tygers} (see \cite{Ray2011}). Those tygers appear shortly before the shock time, at a time $t_{G} \sim t_{s}-K^{-\frac{2}{3}}$ (see eq.~(B5) in \cite{Ray2011}).
    \item When the Galerkin truncated Burgers equation is taken with the minus sign in front of the nonliearity, as it happens for the field $\lambda$ in eq.~\eqref{eq:Burgerissime}, is is known that the tygers appear in the descending profile of $u$ (see Fig.~1 in \cite{Ray2011}, Fig.~1 and the discussion at page 399 in \cite{Ray2015}).
    \item At late times (estimeed in \cite{Venkataraman2017}) the Galerkin truncated Burgers equation exhibits thermalization (see e.g.~Fig~1 in \cite{Ray2015}).
\end{itemize}
Given this short account on the Galerkin truncated Burgers equation, let us briefly discuss the differences and analogy with respect to the system of inviscid Burgers equations \ref{eq:Burgerissime} obtained for the renormalized fields $\lambda,\rho$ in FPUT.

Indeed, the Galerkin truncated Burgers equation is the first order normal form of the FPUT model (compare e.g.~eq.~(29) in \cite{Benettin2008-1D} and eq.~(5) in \cite{Majda2000}. Being the normal form, up to the reminder, means explicitly that the FPUT equations of motion can be written as
\begin{equation}\label{eq:DoppiaKDV}
    \begin{split}
        \lambda_\tau &= \left(1+\frac{\alpha \sqrt{\epsilon}}{\sqrt{2}} \lambda \right) \lambda_x + \mathcal{R}(\lambda,\rho) \\
        \rho_\tau&=-\left(1+\frac{\alpha \sqrt{\epsilon}}{\sqrt{2}} \rho\right)\rho_x - \mathcal{R}(\rho,\lambda)
    \end{split}
\end{equation}
where, for $\epsilon$ small, the leading terms in $\mathcal{R}$ yields the pair of counterpropagating Korteweg-de Vries equations (see e.g.~Eq.~(93) in \cite{Gallone2022} or, more explicitly, equation (6.69) in \cite{Gallone-Review-2025} (The coefficient in front of $\lambda_{xxx}$ is correct in \eqref{eq:DoppiaKDV})
\begin{equation}
    \begin{split}
        \lambda_\tau &= \left(1+\frac{\alpha \sqrt{\epsilon}}{\sqrt{2}} \lambda \right) \lambda_x +\frac{1}{24 \,N^2} \lambda_{xxx}+\widetilde{\mathcal{R}}(\lambda,\rho) \\
        \rho_\tau&=-\left(1+\frac{\alpha \sqrt{\epsilon}}{\sqrt{2}} \rho\right)\rho_x -\frac{1}{24 \,N^2} \rho_{xxx} - \widetilde{\mathcal{R}}(\rho,\lambda)
    \end{split}
\end{equation}
where, for any finite lattice, the regularisation observed is due to the small (but finite, when $N$ is finite) dispersion. Therefore, the dynamics of FPUT -- at the level of equations -- is deeply different from the one of the Galerkin truncated Burgers equation \emph{close} and \emph{after} the shock time. The short wavelength oscillations observed in the FPUT system are due to the small dispersion and, in fact, behave also qualitatively differently from the tygers (see, for example, Fig.~\ref{fig:SpatialPlot}).

\begin{figure}[h!]
    \begin{center}
        \includegraphics[width=0.6\textwidth]{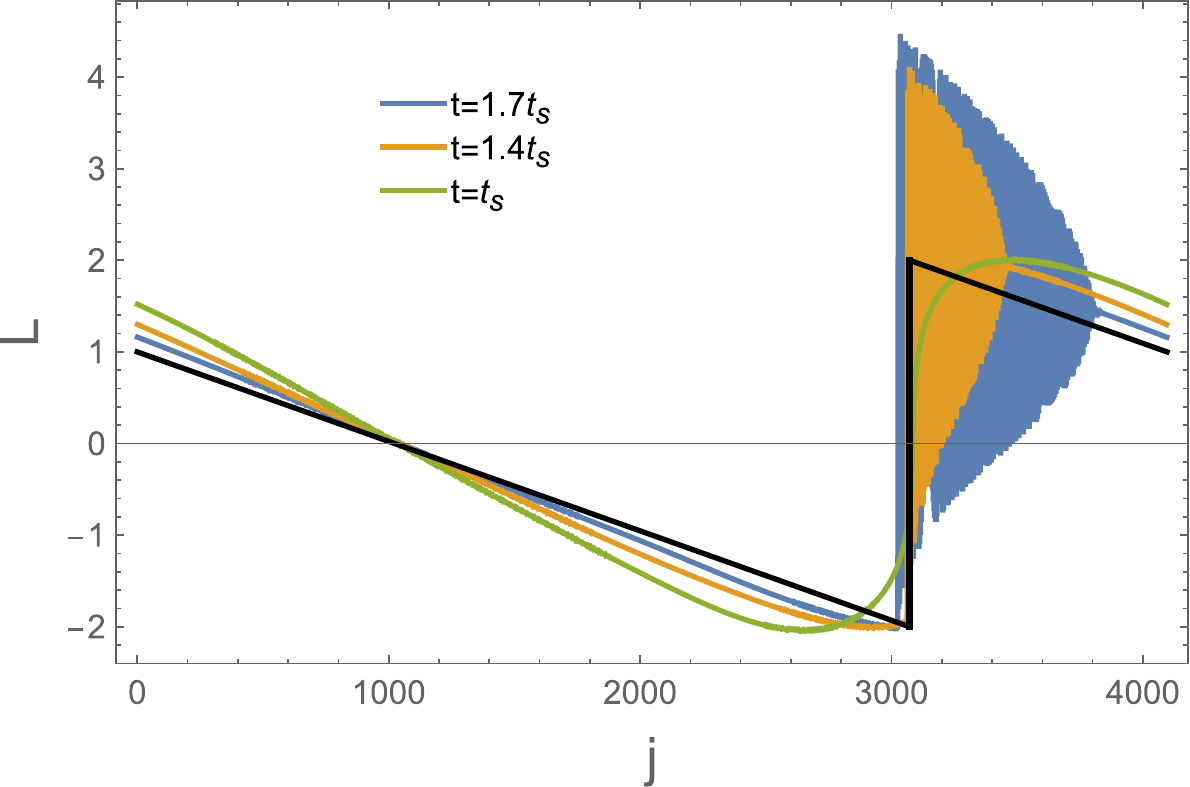}
        \caption{Space profile of the solution of the FPUT system at the shock time and after the shock time. $N=4096$, initial solution with only one mode initially excited with $R=0$ at $t=0$.}\label{fig:SpatialPlot}
    \end{center}
\end{figure}
Indeed, differently from the tygers which would have appeared on the descending part of the space profile, our short wavelength oscillations appear close to the point of the gradient catastrophe.

Having clarified the differences among the two phenomena we would like also to stress similarities, in the \emph{thermalization} phenomenon. For small specific energy, the approach to thermal equilibrium suggested by the truncated Burgers equation is different from the one of the FPUT (this is noted already in \cite{Ray2011}, where the authors write: ``It must be noted that the mechanism which prevents thermalization in the Fermi-Pasta-Ulam problem does not seem to be present here''). Indeed, the route to thermalization must take into account both the dispersion term and all \emph{nonintegrable} terms present in $\widetilde{\mathcal{R}}$.

Nevertheless, when the specific energy is large enough, the thermalization of the FPUT chain is qualitatively similar to the one of the truncated Burgers equation. Indeed, after the shock time, the Fourier Energy Spectrum of the FPUT chain exhibits a stable power law $E_k\sim k^{-2}$ in an inertial range that deteriorates with time (see Fig.~2 and 6 in \cite{Gallone2024}). Correspondingly, in the space profile of the wave, the bunch of short wavelength oscillations formed close to the shock expands until it covers the whole profile. This phenomenon, which is qualitatively described in the concluding part of \cite{Gallone2022-PRL}, is qualitatively similar to what happens with tygers (see e.g.~Fig.~1 in \cite{Avellaneda1995-2}).

\section{Generation of tygers for many-modes initial data}
As a corollary of our main result, we show how the choice of phases affects the time of formation of tygers in the truncated Burgers equation. The argument for a single mode solution can be found in Appendix B of \cite{Ray2011} and, for two modes solution, in Section IV.B of \cite{Rampf2022}.

It is well-known that the solution to the initial value problem of the inviscid Burgers equation
\begin{equation}\label{eq:BurgersAPP}
    \left\{\begin{array}{l}
        u_t(t,x)=u(t,x) u_x(t,x) \\
        u(0,x)=u_0(x)
        \end{array}
    \right.
\end{equation}
for $x \in \mathbb{T}$
exists, as a function in a strong sense, only for $0\leq t \leq t_s$. A lot of research effort has been spent in the Galerkin-truncated Burgers equation
\begin{equation}
    u_t(t,x)=P_K(u(t,x) u_x(t,x)) \, .
\end{equation}
If the initial datum is analytic, any solution to \eqref{eq:BurgersAPP} is analytic until the shock time. In particular, if the initial datum is a Fourier polynomial, the solution to the Burgers equation is analytic on a strip of width $\delta(t)$. This implies, in particular, that
\begin{equation}\label{eq:AnaliticityBurgers}
    |\hat{u}_k(t)| \lesssim e^{-\delta(t) |k|} \, .
\end{equation}
Due to the fact that the cutoff introduced by the Galerkin truncation $P_K$ projects on Fourier modes with $|k| \leq K$, and due to \eqref{eq:AnaliticityBurgers}, it follows that the difference between the solution to the Galerkin truncated Burgers equation and the solution to the inviscid Burgers equation are undistinguishable until $\delta(t) K \sim 1$, that is until the time the poles of the solution intersects the horizontal line in the complex plane $\pm i K^{-1}$. This latter time coincides with the time of formation of the tygers (this is -- in short -- the argument of \cite{Ray2011}).

It is also well-known, both theoretically and numerically, that such a time -- for one-mode solution -- behaves as
\begin{equation}
    t_G=t_s-K^{-\frac{2}{3}} \, .
\end{equation}
In this section, we analyze the tygers formation time for the class of initial solutions we used for FPUT.

The first thing to do, is to compute $t_G$ for generic solutions. This can be done by generalizing the argument of Appendix B in \cite{Ray2011} or following \cite{Rampf2022}. 

The position of the pole in $\xi \in \mathbb{C}$ coordinates is given by the vanishing of the Jacobian of the transformation $x=\xi-u_0(\xi) t$ with $x \in \mathbb{C}$ and $t \in \mathbb{R}$. The condition 
\begin{equation}
    \partial_\xi x = 0 \qquad \Leftrightarrow \qquad 1 - u'_0(\xi) t = 0
\end{equation}
Calling now $\xi_* \in \mathbb{R}$ the point where the first pole hits the real line, by definition of the shock time we have $t_s=\frac{1}{u'_0(\xi_*)}$, which yields also $1=u'_0(\xi_*) t_s$. By expanding $u_0'(\xi)$ around $\xi_*$ we have
\begin{equation}
    u'_0(\xi_*) t_s-u'_0(\xi_*) t-\frac{1}{2}u_0'''(\xi_*)(\xi-\xi_*)^2 t_s-\frac{1}{2}u_0'''(\xi_*)(\xi-\xi_*)^2(t-t_s)+\cdots = 0
\end{equation}
where we assumed a genericity condition $u_0'''(\xi_*) \neq 0$.
Using now the fact that $t-t_s$ is small, we obtain, up to higher order terms, the equation to solve in $\xi$
\begin{equation}
    u'_0(\xi_*)(t_s-t)-\frac{1}{2}u_0'''(\xi_*)(\xi-\xi_*)^2 t_s = 0 \, .
\end{equation}
The latter yields
\begin{equation}\label{eq:PoliXi}
    \xi-\xi_*= \pm \sqrt{\frac{2(t_s-t)u_0'(\xi_*)}{u_0'''(\xi_*)t_s}}=\pm \frac{i}{t_s}\sqrt{\frac{2(t_s-t)}{|u_0'''(\xi_*)| }}
\end{equation}
where in the last step we used the fact that, since $\xi_*$ is a maximum for the first derivative, then $(u_0')''(\xi_*) \leq 0$.

We now insert \eqref{eq:PoliXi} into $x=\xi-u_0(\xi) t$ to compute the poles in the $x$ variables. The formation of the tyger is related to the time where the poles are at $\Im x=\pm i K^{-1}$. Thus,
\begin{equation}
    \begin{split}
    x_*&=\xi-\xi_*+\xi_* -u_0(\xi_*) t -u'_0(\xi_*)(\xi-\xi_*)t - \frac{1}{6}u_0'''(\xi_*)(\xi-\xi_*)^3 t_s + \cdots \\
    &=\xi_*- u_0(\xi_*) t +u'_0(\xi_*)(\xi-\xi_*)(t_s-t)-\frac{1}{6} u_0'''(\xi_*)(\xi-\xi_*)^3t_s+ \cdots \\
    &=\xi_*- u_0(\xi_*) t +\frac{1}{t_s}(\xi-\xi_*)(t_s-t)-\frac{1}{6} u_0'''(\xi_*)(\xi-\xi_*)^3t_s+ \cdots
    \end{split}
\end{equation}
Taking now the imaginary part at both sides, one gets
\begin{equation}
    \begin{split}
        K^{-1}&=\frac{1}{t_s^2} \sqrt{\frac{2}{|u_0'''(\xi_*)|}}(t_s-t)^{\frac{3}{2}}-\frac{1}{3 t_s^2} \sqrt{\frac{2}{|u_0'''(\xi_*)|}}(t_s-t)^{\frac{3}{2}}+ \cdots \\
        &=\frac{2 \sqrt{2}}{3 t_s^2} \frac{1}{\sqrt{|u_0'''(\xi_*)|}} (t_s-t)^{\frac{3}{2}} \, .
    \end{split}
\end{equation}
which yields
\begin{equation}
    t_s-t=\left(\frac{3 }{2\sqrt{2}}  \right)^{\frac{2}{3}} |u_0'''(\xi_*)|^{\frac{1}{3}} t_s^{\frac{4}{3}}K^{-\frac{2}{3}}
\end{equation}
This implies
\begin{equation}
    t_G=t_s\left[ 1-\left(\frac{3}{2 \sqrt{2}}\right)^{\frac{2}{3}} |u_0'''(\xi_*)|^{\frac{1}{3}} t_s^{\frac{1}{3}} K^{-\frac{2}{3}}\right]
\end{equation}
which agrees with Eq.~(B5) in \cite{Ray2011} when the initial datum is a sine-wave. Let us now check what happens for the initial data of our problem, that is when
\begin{equation}
    u(x)=\frac{1}{\sqrt{p}} \sum_{k=1}^p  \cos(2 \pi k x + \theta_k^\omega)
\end{equation}
When all $\theta_k^\omega=0$, that is the case of coherent phases, we obtain
\begin{equation}
    t_s^{-1}=u'(0)=\frac{1}{\sqrt{p}}\sum_{k=1}^p (2 \pi k)=\sqrt{p}(p+1) \pi \, , \qquad u_0'''(0)=-\frac{1}{\sqrt{p}} \sum_{k=1}^p(2 \pi k)^3=-2 p^{\frac{3}{2}}(1+p)^2 \pi^3
\end{equation}
therefore, one obtains as the time of formation of the first tyger the explicit expression
\begin{equation}
    t_G=\frac{1}{\sqrt{p}(p+1) \pi} \left[1-\left(\frac{3}{2} \right)^{\frac{2}{3}} \pi \sqrt{p}(1+p)^{\frac{2}{3}} K^{-\frac{2}{3}} \right]
\end{equation}
In particular, the last equation says that, with respect to the shock time, one has
\begin{equation}
    1-\frac{t_G}{t_s}=\left(\frac{3}{2}\right)^{\frac{2}{3}} \pi \sqrt{p}(1+p)^{\frac{2}{3}}K^{-\frac{2}{3}} \, .
\end{equation}

When random initial data are taken, in general the behavior is much different. Indeed, on average, one has the following upper bound
\begin{equation}
    1-\mathbb{E}\left[ \frac{t_G}{t_s}\right] =\left(\frac{3}{2\sqrt{2}}\right)^{\frac{2}{3}} \mathbb{E}\left[|u_0'''(\xi_*)|^{\frac{1}{3}} t_s^{\frac{1}{3}}\right] K^{-\frac{2}{3}}
\end{equation}
Using that $t_s \leq \frac{1}{p}$, that is because 
\begin{equation}
    t_s = \frac{1}{\sup_\xi u_0'(\xi)} \lesssim \frac{1}{p}
\end{equation}
because $\sup_\xi (u_0')^2 \geq \int_0^1 u_0'(\xi)^2 \, d\xi = \frac{1}{p\sqrt{2}}\sum_{k=1}^p  (2 \pi k)^2$ where in the last step we used the Parseval identity.

we get
\begin{equation}
    \mathbb{E}\left[|u_0'''(\xi_*)|^{\frac{1}{3}} \right] p^{-\frac{1}{3}} \leq \mathbb{E}\Big[ \sup_{\xi} |u_0'''(\xi)|^{\frac{1}{3}} \Big] p^{-\frac{1}{3}}
\end{equation}
Using H\"older inequality, we have $\mathbb{E}\Big[ \sup_{\xi} |u_0'''(\xi)|^{\frac{1}{3}}\Big] \leq \mathbb{E}\left[ \sup_{\xi} u_0'''(\xi) \right]^{\frac{1}{3}}$. One can now estimate $\mathbb{E}\left[ \sup_\xi u_0'''(\xi) \right]$ as we did for $\mathbb{E}[\sup_{\xi}u_0'(\xi)]$ above, to get
\begin{equation}
    \mathbb{E}\Big[ \sup_\xi u_0'''(\xi) \Big]^{\frac{1}{3}} \lesssim p^{\frac{7}{6}} (\log p)^{\frac{1}{6}} \, .
\end{equation}
We thus obtained 
\begin{equation}
   0\leq 1-\mathbb{E}\left[\frac{t_G}{t_s} \right] \lesssim p^{\frac{5}{6}} (\log p)^{\frac{1}{6}} \, K^{-\frac{2}{3}} \, .
\end{equation}
with a constant independent of $p$ and $K$. In particular, the right-hand side is small whenever $K \gtrsim p^{\frac{5}{4}} (\log p)^{\frac{1}{4}}$.


\section{Some details on the numerics}
In this Section, we present an example of FES and explain how the shock time is computed numerically.

We integrate the FPUT system using a fourth-order Yoshida algorithm. Since this algorithm is symplectic, the total energy is conserved with high accuracy (relative error $\lesssim 10^{-5}$).

To identify the shock time, we monitor the FES on a log–log scale at successive times. At the shock time, the spectrum develops an inertial range with scaling $E_k \sim k^{-8/3}$ and a sharp  ``knee'' at its right edge. Before the shock time this knee is rounded, while after it evolves into a hump that collects the energy of short-wavelength solitons generated by the lattice dispersion.

\begin{figure}[h!]
    \includegraphics[width=0.49\textwidth]{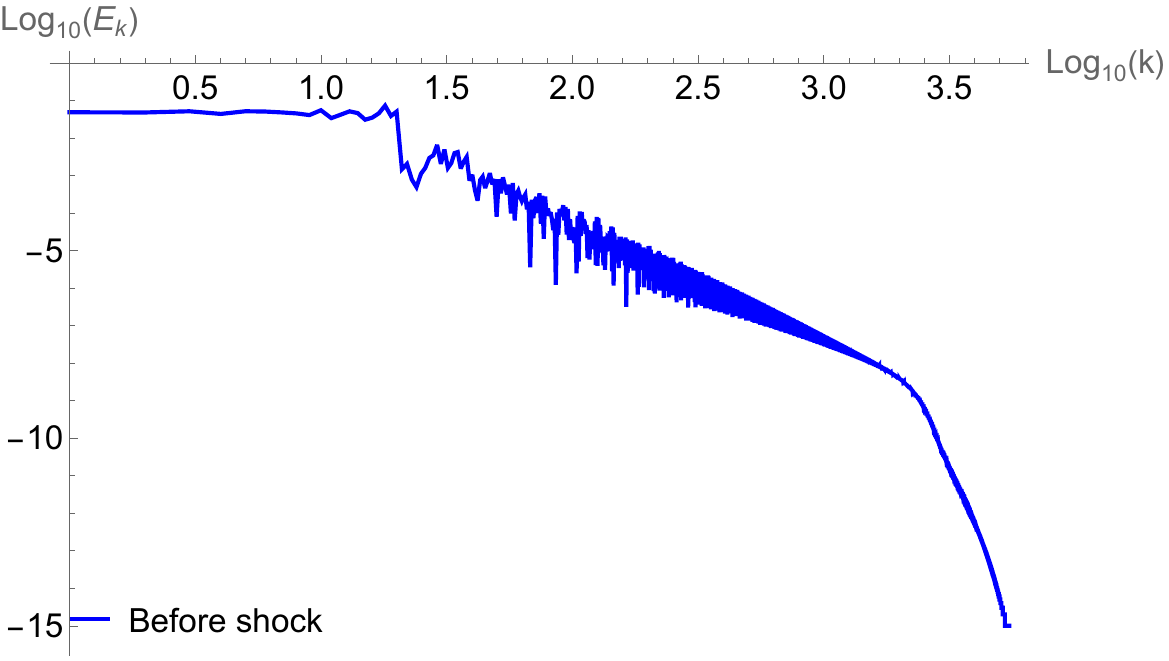} \includegraphics[width=0.49\textwidth]{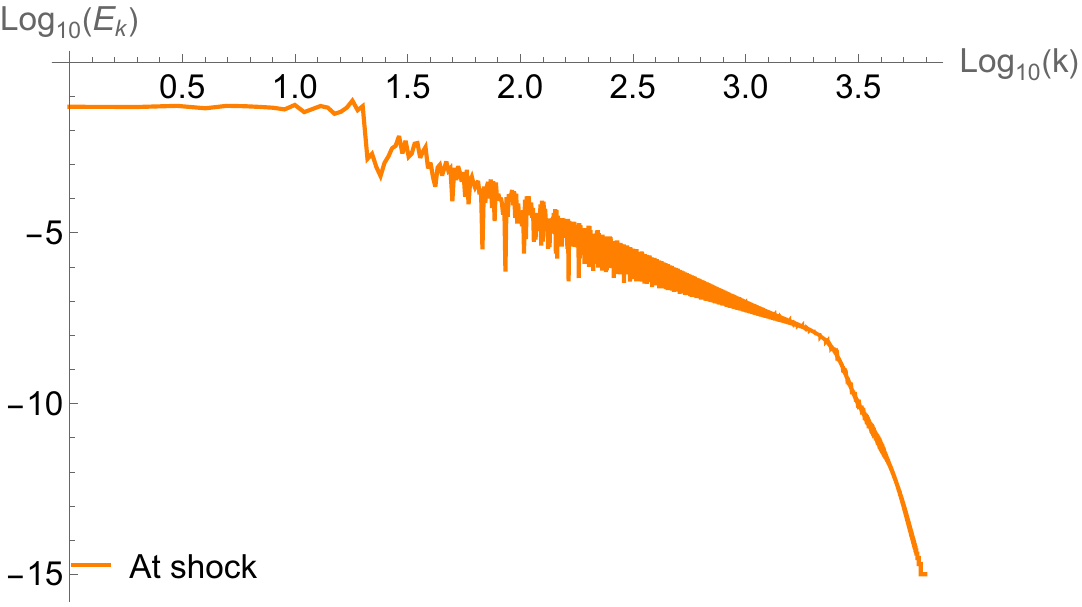}

    \includegraphics[width=0.49\textwidth]{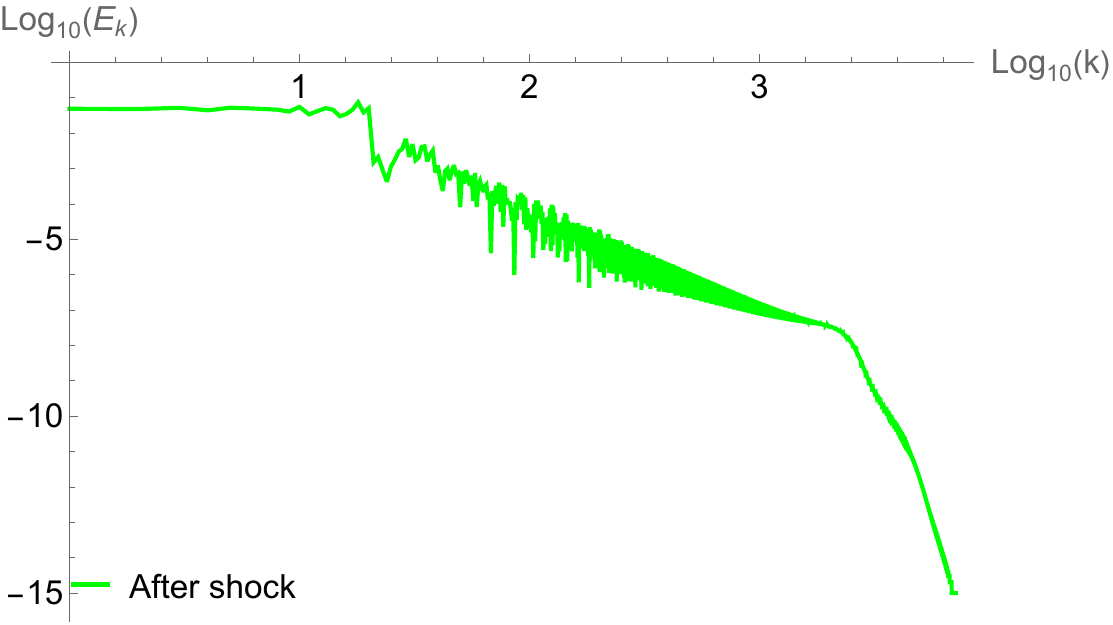} \includegraphics[width=0.49\textwidth]{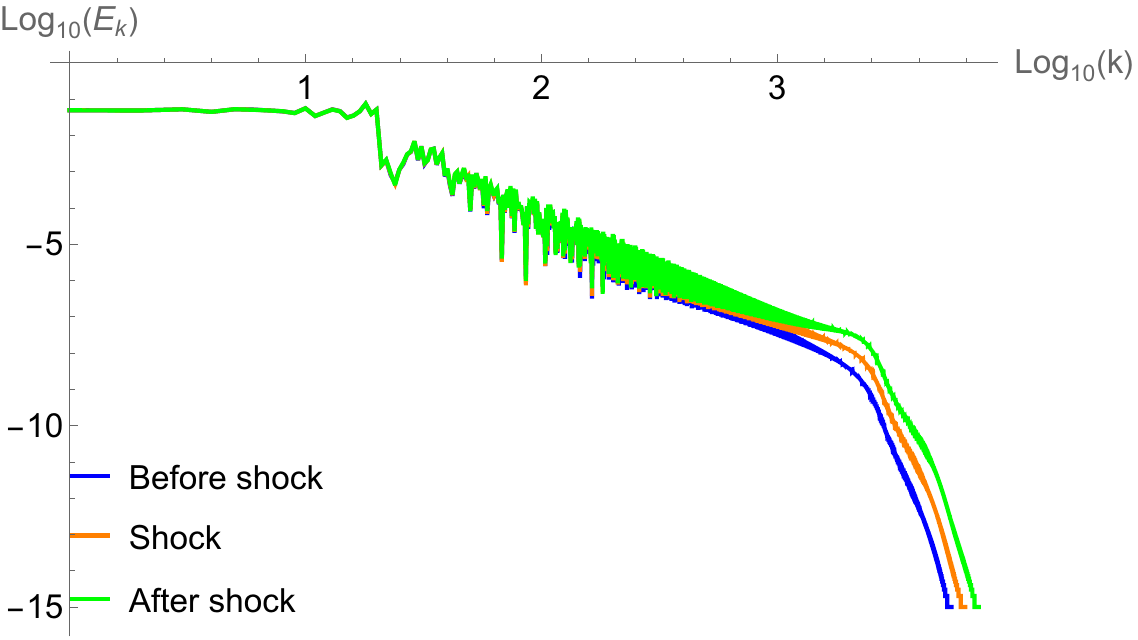}

    \caption{Fourier Energy spectrum of the FPUT chain with random initial data in a short interval around the shock time. In the first panel, the blue line represents the FES at $\sim 97\%$ shock time; in the second panel it is represented the FES at the shock time and in the third panel the FES at $\sim 103\%$ shock time. In the last panel we superimpose the three FES.}
\end{figure}

To better highlight the power-law scaling, we reduce noise by computing the maximum of $E_k$ over windows in $k$ of width 10, and refer to the resulting curve as the ``filtered spectrum''. This filtering step is essential to determine the shock time accurately.

\begin{figure}[h!]
    \begin{center}
    \includegraphics[width=0.5\textwidth]{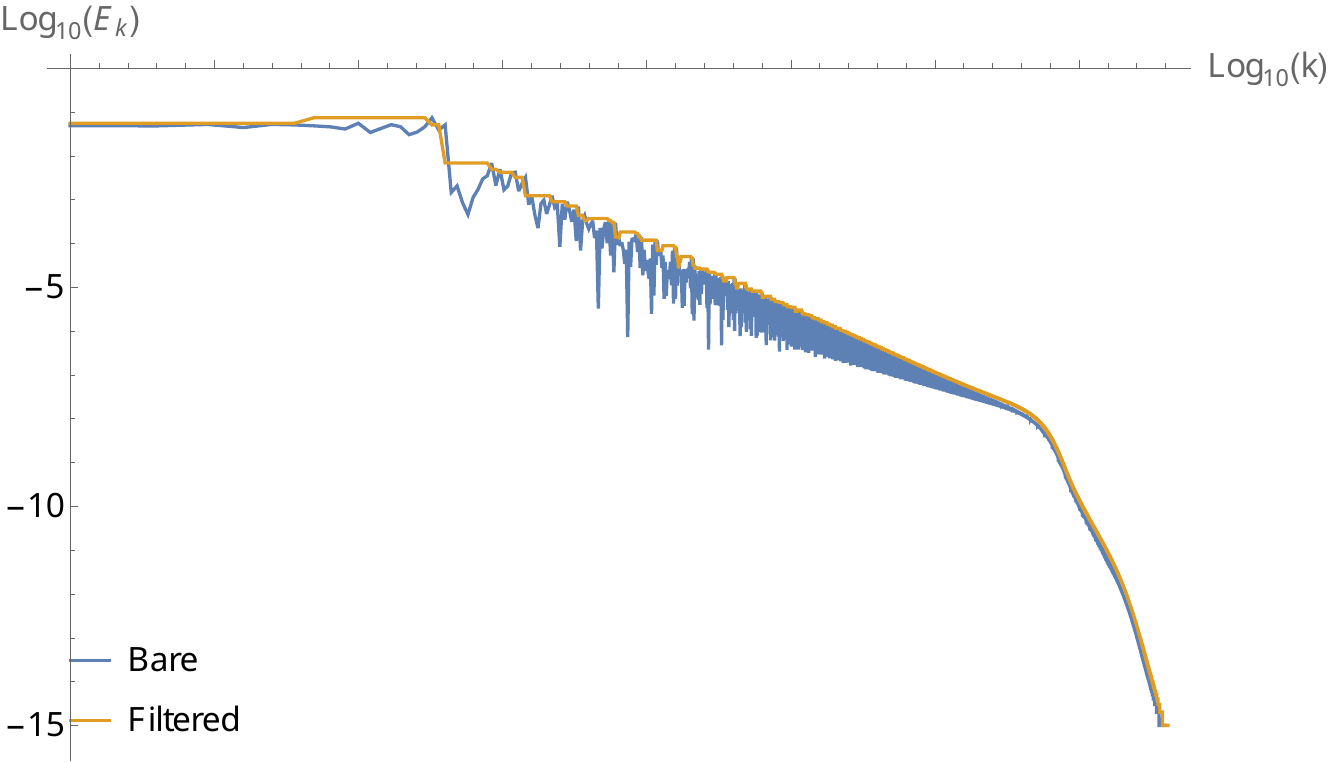}
    \caption{Filtered and bare FES at the shock time.}
    \end{center}
\end{figure}

 For each data point in Fig.~\ref{FigLog}, we sampled $1000$ random initial conditions for a system with $2^{14}=16\,384$ masses. We let the system evolve and on a certain time-window (which depends on the number of excited modes) we computed the FES at any time and we plotted it. We then estimated the shock time using the procedure described above. The plotted values are the averages over these realizations at fixed $p$, the number of initially excited modes.

Random phases are generated using standard C++ functions. The code is:
\begin{lstlisting}
double random_0_to_2pi() {
    return (static_cast<double>(rand()) / RAND_MAX) * (2 * M_PI);
}
    
\end{lstlisting}

\end{widetext}

\end{document}